\documentclass[journal]{IEEEtran}

\usepackage{xcolor}
\usepackage[normalem]{ulem}
\usepackage{probsoln}
\usepackage{array}
\usepackage[most]{tcolorbox}

\usepackage{xparse}
\usepackage{wrapfig}
\usepackage{booktabs}
\usepackage[colorlinks]{hyperref}
\usepackage{cite}

\hypersetup{
	colorlinks=true, 
	linkcolor=blue!70!black, 
	citecolor=red!70!black, 
	filecolor=blue!70!black, 
	urlcolor=blue!70!black, 
}

\newcommand{\J}{\mathrm{j}}                 
\newcommand{\degree}{\ensuremath{^\circ}}

\providecommand{\T}[1]{\mathrm{#1}}         

\providecommand{\ZVAC}{\eta_0}  
\providecommand{\Zmatch}{Z_0}  
\providecommand{\Ohm}{\,\Omega}

\providecommand{\Wsto}{W_\T{sto}}

\providecommand{\Rin}{R_\T{in}}
\providecommand{\Xin}{X_\T{in}}

\providecommand{\RL}{R_\T{L}}
\providecommand{\XL}{X_\T{L}}
\providecommand{\Prad}{P_\T{rad}}
\providecommand{\Plost}{P_\T{diss}}
\providecommand{\Rsurf}{R_\T{s}}
\providecommand{\BW}{B}

\providecommand{\Qrad}{Q_\T{rad}} 
\providecommand{\Qradmin}{\Qrad^\T{lb}} 
\providecommand{\QradminTM}{\Qrad^\T{lb,TM}} 
\providecommand{\DissipF}{\delta}
\newcommand{\meanderline}{meander line}

\newcommand{\ie}{\textit{i.e.}{}}
\newcommand{\eg}{\textit{e.g.}{}}
\newcommand{\cf}{\textit{cf.}{}}
\newcommand{\Quot}[1]{``{#1}"} 

\newcommand\figwidth{9cm}
\graphicspath{{figures/}}
\newtcolorbox[auto counter]{boxexampl}[2][]{floatplacement=t,float,%
colback=blue!0!white,colframe=blue!75!black,fonttitle=\bfseries,fontupper=\small,title=Box~\thetcbcounter. #2,#1}

\begin{document}
\title{Optimal Planar Electric Dipole Antennas}

\author{
Miloslav~Capek,~\IEEEmembership{Senior~Member,~IEEE}, 
Lukas~Jelinek,
Kurt~Schab,~\IEEEmembership{Member,~IEEE}, 
Mats~Gustafsson,~\IEEEmembership{Senior~Member,~IEEE}, 
B.\,L.\,G.~Jonsson,
Fabien~Ferrero,
and Casimir~Ehrenborg,~\IEEEmembership{Student Member,~IEEE}
\thanks{M.~Capek and L.~Jelinek are with the Department of Electromagnetic Field, Faculty of Electrical Engineering, Czech Technical University in Prague, Technicka~2, 16627, Prague, Czech Republic
(e-mail: \{miloslav.capek, lukas.jelinek\}@fel.cvut.cz).}%
\thanks{K.~Schab is with the Electrical Engineering Department, Santa Clara University, 500~El Camino Real, Santa Clara, CA~95053 (e-mail: kschab@scu.edu).}
\thanks{M. Gustafsson and C. Ehrenborg are with the Department of Electrical and Information Technology, Lund University, Box 118, SE-221 00 Lund, Sweden. (Email: \{mats.gustafsson, casimir.ehrenborg\}@eit.lth.se).}%
\thanks{B.~L.~G.~Jonsson is with the School of Electrical Engineering and Computer Science, KTH Royal Institute of Technology, 100 44 Stockholm, Sweden. (Email: ljonsson@kth.se).}%
\thanks{F.~Ferrero is with the Laboratory of Electronics, Antennas and Telecommunications, CNRS, Universit\'{e} C\^{o}te d'Azur, 06903 Sophia Antipolis,
France. (e-mail: fabien.ferrero@unice.fr).}	
\thanks{This work was supported by the Swedish Foundation for Strategic Research (SSF), grant no.~\mbox{AM13-0011}, by VINNOVA through ChaseOn/iAA and by the Czech Science Foundation under project \mbox{No.~19-06049S}. The work of M.~Capek was supported by the Ministry of Education, Youth and Sports through the project CZ.02.2.69/0.0/0.0/16\_027/0008465.}
}


\maketitle

\begin{abstract}
Considerable time is often spent optimizing antennas to meet specific design metrics.  Rarely, however, are the resulting antenna designs compared to rigorous physical bounds on those metrics. Here we study the performance of optimized planar \meanderline~antennas with respect to such bounds.  Results show that these simple structures meet the lower bound on radiation Q-factor (maximizing single resonance fractional bandwidth), but are far from reaching the associated physical bounds on efficiency.  The relative performance of other canonical antenna designs is compared in similar ways, and the quantitative results are connected to intuitions from small antenna design, physical bounds, and matching network design.
\end{abstract}


\IEEEpeerreviewmaketitle

\section{Introduction}
\IEEEPARstart{A}{ntenna} parameters such as gain, Q-factor, and efficiency are limited by the geometry made available for a given design. Given bounds on these parameters under certain constraints, a designer can rapidly assess the feasibility of design requirements.  This feasibility assumes the existence of an \Quot{optimal antenna} design which approaches the bounds on certain specified parameters.  Synthesis of an optimal antenna is not a trivial task, and it remains to be demonstrated how an antenna designed to be optimal in one parameter (\eg{}, radiation Q-factor) performs relative to bounds on other parameters (\eg{}, efficiency).  The goal of this paper is to discuss the synthesis and analysis of optimal antennas starting from classical antenna topologies.

Many strategies have been employed to optimize antennas.  Heuristic optimization methods such as genetic algorithms~\cite{RahmatMichielssen_ElectromagneticOptimizationByGenetirAlgorithms,Haupt+Werner2007} and particle swarm optimization have the advantage of generating design geometries outside of the antenna designer's usual catalog \cite{RahmatSamii_Kovitz_Rajagopalan-NatureInspiredOptimizationTechniques, OnwuboluBabu_NewOptimizationTechniquesInEngineering, Deb_MultiOOusingEA}. Such techniques have been used to design optimal antennas with radiation Q-factors very close to the physical bounds~\cite{CismasuGustafsson_FBWbySimpleFreuqSimulation}, though the resulting designs are computationally expensive to produce and offer only rough insight into guidelines for designing optimal antennas in volumes with arbitrary shapes and electrical size. Conversely, canonical antenna designs were shown~\cite{GustafssonSohlKristensson_IllustrationsOfNewPhysicalBoundOnLinearlyPolAntennas, Best_ElectricallySmallResonantPlanarAntennas} to reach the lower bound on radiation Q-factor, but the question remains whether these designs represent optimal solutions over arbitrary electrical sizes and whether they are optimal in other parameters, \eg{}, radiation efficiency and input impedance. The cost of matching an optimal antenna design to arbitrary impedances is also unclear, regardless if matching is performed on the antenna itself or through external networks.

In this paper we study whether there exists a simple \Quot{recipe} for an optimal planar antenna (Throughout this paper, the term planar means to lie in a plane.) with respect to radiation Q-factor and radiation efficiency.  In doing so, we ask whether, when prescribed with some form factor and electrical size, a simple design can be readily employed to achieve an antenna whose properties are sufficiently close to their bounds. The strategy adopted here is to optimize parameterizations of canonical antenna geometries known for good behavior in certain parameters.  The examples studied here give quantifiable results to this end, \ie{}, how to design certain kinds of optimal antennas.

Along the way, we address the crossover of optimality of antennas across different performance parameters, \eg{}, do minimum radiation Q-factor antennas have inherently high radiation efficiency?  Also discussed are the impacts of certain constraints, particularly those related to an antenna's input impedance, on optimized parameters.

We stress out that this work differs significantly from other works on antenna optimization through parametric, heuristic, or metaheuristic means which typically involves the iterative evaluation and modification of designs until a local optimum or design goal is reached. Here, instead, we focus on designing antenna performance with respect to physical bounds, which provide an absolute measure in judging the quality of the synthesized design.

\section{Minimum Radiation Q-factor of planar TM antennas}
\label{sec:Q-optimization}

We begin by studying the synthesis of electrically small dipole-like (TM) antennas with minimal radiation Q-factor~$\Qrad$~(see Box~\ref{B:Box1a} and Box~\ref{B:Box1b}). This leads to increased impedance bandwidth, however, the lower bound on radiation Q-factor increases rapidly as an antenna design region becomes smaller (see Box~\ref{B:Box1b}).  Thus, obtaining low Q-factor~$\Qrad$ is a key objective and challenge in the design of electrically small antennas.

\begin{boxexampl}[label={B:Box1a}]{{Q-factor}}
The Q-factor of an antenna system tuned to resonance (equality of mean magnetic and mean electric energy) is defined as~\cite{IEEEStd_antennas}
\begin{equation}
Q = \frac{2\omega \Wsto}{\Prad + \Plost},
\label{eq:qdef_box1a}
\end{equation}
where~$\Wsto$ represents the cycle mean stored energy, while~$\Prad$ and~$\Plost$ denote cycle mean radiated and dissipated powers, respectively.  In single-resonance systems, lower Q-factor implies larger fractional impedance bandwidth~$\BW$ by an inverse relationship~\cite{Chu_PhysicalLimitationsOfOmniDirectAntennas,Fante1969,YaghjianBest_ImpedanceBandwidthAndQOfAntennas}
\begin{equation}
\BW \sim Q^{-1}.
\label{eq:b_q}
\end{equation}

Evaluated at a single frequency via~\eqref{eq:qdef_box1a}, Q-factor thus becomes a convenient measure of the frequency selectivity of a system~\cite{Chu_PhysicalLimitationsOfOmniDirectAntennas,Collin+Rothschild1964,Fante1969,Harrington+Mautz1972,Rhodes1976,Collin1998,YaghjianBest_ImpedanceBandwidthAndQOfAntennas,Vandenbosch_ReactiveEnergiesImpedanceAndQFactorOfRadiatingStructures,GustafssonCismasuJonsson_PhysicalBoundsAndOptimalCurrentsOnAntennas_TAP,Geyi2011,Gustafsson+Jonsson2015b,CapekGustafssonSchab_MinimizationOfAntennaQualityFactor,Gustafsson+Jonsson2015a}.  Calculation of a system's Q-factor can be carried out by a variety of approaches, from impedance-based techniques~\cite{YaghjianBest_ImpedanceBandwidthAndQOfAntennas} to methods based on the evaluation of stored energy directly~\cite{Vandenbosch_ReactiveEnergiesImpedanceAndQFactorOfRadiatingStructures, CismasuGustafsson_FBWbySimpleFreuqSimulation}.  All of these approaches generally agree for electrically small, narrow-band antennas, see~\cite{Schab+etal2018} for complete discussion and bibliography. \\

The radiation Q-factor~$\Qrad$, in which only radiated power is considered, can be expressed in terms of $Q$ in \eqref{eq:qdef_box1a} and radiation efficiency~$\eta$ (see Box~\ref{B:delta1}, \eqref{Box_eta}) as 
\begin{equation}
\Qrad = Q / \eta \sim (\BW\eta)^{-1}.
\label{eq:b_eta}
\end{equation}
\end{boxexampl}

\begin{boxexampl}[label={B:Box1b}]{{Lower bounds on radiation Q-factor}}
The approximate inverse proportionality between the \mbox{Q-factor} and the fractional bandwidth (see Box~\ref{B:Box1a}) induced considerable effort in lowering the Q-factor for spherical~\cite{Chu_PhysicalLimitationsOfOmniDirectAntennas,VolakisChenFujimoto_SmallAntennas,Thal2006,Collin+Rothschild1964,McLean1996,Kim2012} and arbitrarily shaped antennas~\cite{Wheeler1947,Gustafsson+etal2015b,Gustafsson+etal2007a,Yaghjian+etal2013,CapekJelinek_OptimalCompositionOfModalCurrentsQ,Sievenpiper+etal2012,Thal2012}. The sole focus on the radiation Q-factor~$\Qrad$ in these works avoids the undesired possibility of reducing Q-factor~$Q$ by degrading radiation efficiency.\\

For electrically small antennas, the lower bound on the~$\Qrad$, here denoted as~$\Qradmin$, is a combination of electric and magnetic dipoles~\cite{Chu_PhysicalLimitationsOfOmniDirectAntennas,McLean1996, Jonsson+Gustafsson2015,CapekJelinek_OptimalCompositionOfModalCurrentsQ,Best2005}. 
In general, it is challenging to excite such a current with a single feeding position see \cite[Sec.~IV]{JelinekCapek_OptimalCurrentsOnArbitrarilyShapedSurfaces} for related discussion. A constrained minimization which is more representative for single port antennas is to restrict the radiation to TM (electric dipole) modes, yielding the lower bound $\QradminTM$, see, \eg, \cite{GustafssonTayliEhrenborgEtAl_AntennaCurrentOptimizationUsingMatlabAndCVX}.

The importance of Q-factor bounds arises from two key properties. First, the Q-factor bound represents the physical lower bound among all possible currents contained within the considered region. It thus presents an absolute measure against which to compare the performance of different antenna designs. Practical feasibility of designing antennas which reach various bounds remains an open question. Second, both $\Qradmin$ and $\QradminTM$ scale approximately as $(ka)^{-3}$ for electrically small antennas ($ka<1$), \cf~\cite{Jonsson+Gustafsson2016}. Here $k$ is the free-space wavenumber and $a$ is the radius of the smallest circumscribing sphere.  This $(ka)^{-3}$ scaling is the root cause of the limited bandwidth in electrically small antennas. The associated geometry coefficients for certain shapes are shown in Table~\ref{Tab:Bounds}, where $\ZVAC$ denotes free-space impedance and $R_\mathrm{s}$ denotes surface resistance.
\end{boxexampl}

\begin{table}[t] 
\centering 
\caption{Lower bounds on radiation Q-factors and efficiency in the limit of electrical size $ka \to 0$ for a sphere, a cylindrical tube, a rectangle, a thin strip dipole and a square loop.}
\begin{tabular}{cccc}		
\rule{0pt}{3ex}\rule[-2ex]{0pt}{0pt} & $(ka)^3 \Qradmin$ & $(ka)^3 \QradminTM$ & $(ka)^4 \ZVAC/R_\mathrm{s} \delta$ \\ \toprule 
\noindent\parbox[c]{2cm}{\centering\includegraphics[scale=1.0]{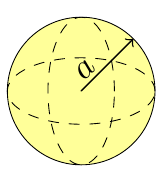}} & 1 & $\displaystyle\frac{3}{2}$ & 3 \\ \midrule 
\noindent\parbox[c]{2cm}{\centering\includegraphics[scale=1.0]{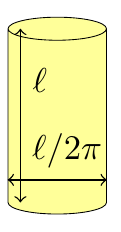}} & 
4.5 & 
4.6 & 
59 \\ \midrule 
\noindent\parbox[c]{2cm}{\centering\includegraphics[scale=1.0]{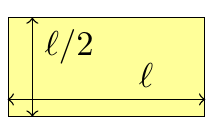}} & 4.3 & 
5.2 & 42 \\ \midrule 
\noindent\parbox[c]{2cm}{\centering\includegraphics[scale=1.0]{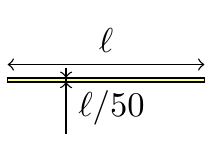}} & 16 & 16 & 3400 \\ \midrule 
\noindent\parbox[c]{2cm}{\centering\includegraphics[scale=1.0]{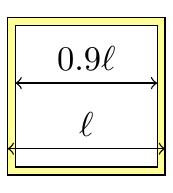}} & 5.3 & 7.4 & 130 \\ \bottomrule 
\end{tabular} 
\label{Tab:Bounds}
\end{table}

\subsection{Synthesis of \meanderline~antennas}
Drawing from the prevalence of \meanderline~antennas in applications requiring electrically small planar antennas \cite{Fujimoto_Morishita_ModernSmallAntennas, VolakisChenFujimoto_SmallAntennas}, as well as previous work studying their optimality in radiation Q-factor~\cite{Best_ElectricallySmallResonantPlanarAntennas}, we focus on determining whether \meanderline s present a consistent, simple solution, to obtaining minimum radiation Q-factor at arbitrary frequencies within rectangular design regions. Here, and throughout Section~\ref{sec:Efficiency-optimization}, we specify a rectangular design region of fixed aspect ratio (\mbox{$L/W = 2$}). The impact of varying aspect ratios is demonstrated and discussed in Section~\ref{sec:q-aspect-ratios}.

\begin{figure}
\centering
\includegraphics[scale=1]{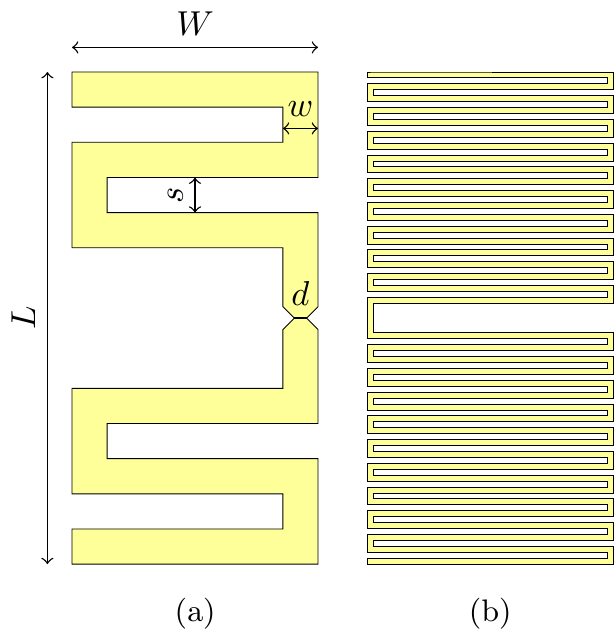}
\caption{Panel~(a) shows a parameterization of a \meanderline~antenna used in this work. The antenna is fed via delta gap source along a horizontal line cutting the center of the \meanderline. The feeding region contains a taper between the feeding strip of width~$d$ and the \meanderline~of width~$w$. The angle of the taper's cut is~$45\degree$. Throughout the paper \mbox{$d=\mathrm{min} \left(w,L/40 \right)$} in order to keep the feeding region realistically narrow. Panel~(b) shows a \meanderline~antenna design \Quot{M1} from~\cite{Best_ElectricallySmallResonantPlanarAntennas} within the parametrization used in this paper.}
\label{fig:meander-parameterization}
\end{figure}

\begin{figure}
\includegraphics[width=\figwidth]{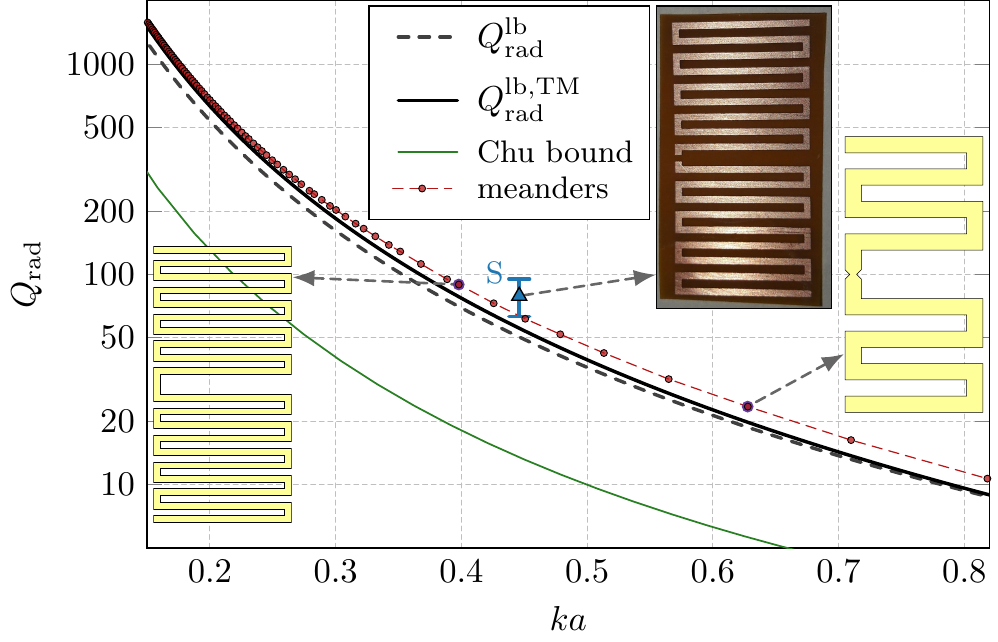}
\caption{Radiation Q-factors of self-resonant \meanderline~antennas simulated as made of PEC (markers) and the lower bound on radiation Q-factor (see Box~\ref{B:Box1a}) corresponding to a rectangular region bounding the \meanderline. All \meanderline s are designed using the parameterization in Fig.~\ref{fig:meander-parameterization} with \mbox{$w/s = 1$} and \mbox{$L/W=2$}. The defining parameters of all meandered dipoles are depicted in Fig.~\ref{fig:meander-designs}. A triangular marker with a corresponding error bar represents measured radiation Q-factor of selected \meanderline~design.}
\label{fig:meander-designs-vs-qbound}
\end{figure}

From the many possible \meanderline~shapes (for example, rectangular, triangular, sinusoidal~\cite{Fujimoto_Morishita_ModernSmallAntennas}) we have chosen the simple parametrization from~Fig.~\ref{fig:meander-parameterization}.  Thin wire versions of such antennas were previously shown to reach the lower bound on radiation Q-factor~$\Qrad$ for their corresponding rectangular design regions with electrical sizes near~\mbox{$ka = 0.3$}~\cite{Best_ElectricallySmallResonantPlanarAntennas}.  Here, we use the parameterization in Fig.~\ref{fig:meander-parameterization} to optimize the \meanderline~antenna for resonance by requiring the magnitude of the normalized input reactance~\mbox{$\Xin/\Rin$} to be smaller than a specified tolerance, \mbox{$\left|\Xin/\Rin \right| < 10^{-3}$}. This procedure is repeated at many frequencies (electrical sizes, values of~$ka$) to obtain a set of antenna designs, each resonant at a specific frequency. The Q-factors~$\Qrad$ of the resulting designs were then calculated in AToM~\cite{atom} and compared to the bounds discussed in Box~\ref{B:Box1b}. The comparison is shown in Fig.~\ref{fig:meander-designs-vs-qbound}. Note that the value of Q-factor~$\Qrad$ is just weakly dependent on dissipation factor (see Box~\ref{B:delta1}) provided that dissipation is not exceedingly high.

\begin{figure}
\includegraphics[width=\figwidth]{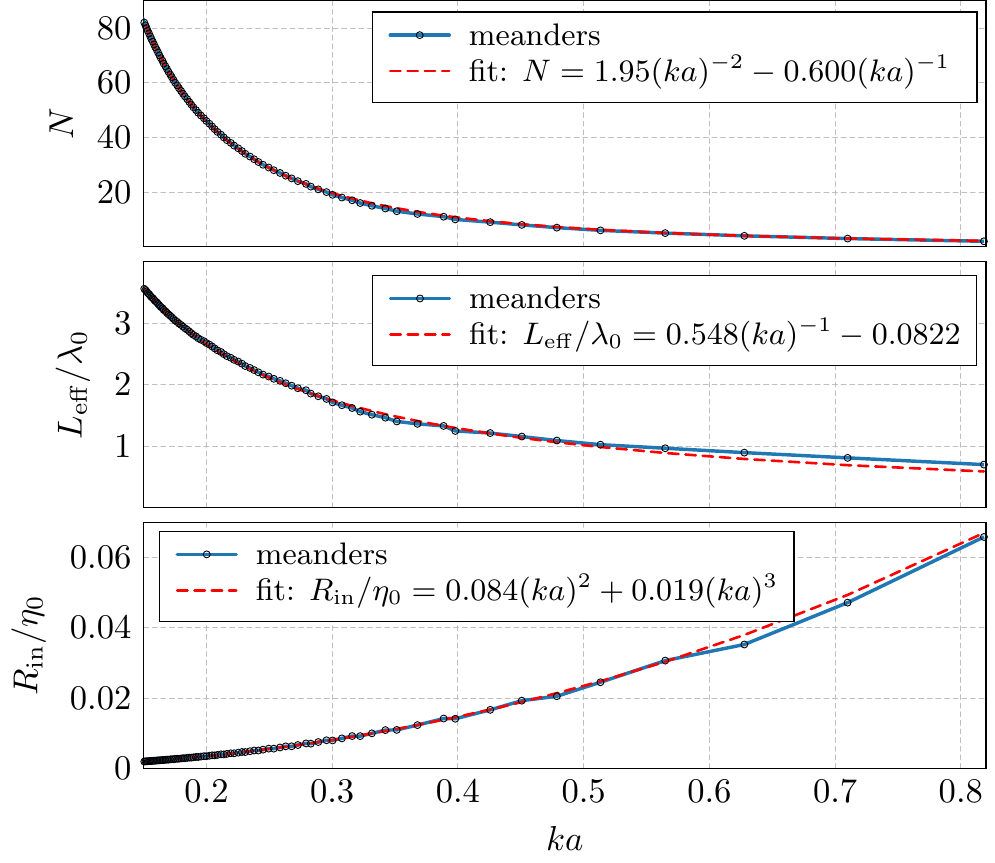}
\caption{Design curves for the \meanderline~antennas from Fig.~\ref{fig:meander-parameterization}. The panels show consecutively total number of meanders~$N$, total length of the strip~$L_\mathrm{eff}$ normalized to free-space wavelength, and input resistance~$R_\mathrm{in}$ normalized to free-space impedance. In all cases~\mbox{$w/s = 1$}, \mbox{$L/W = 2$} and PEC are considered. One \meanderline~(\mbox{$N=1$}) consists of two horizontal strips and vertical connections, \ie{}, \meanderline~antennas in Fig.~\ref{fig:meander-parameterization} have 3~and 20~meanders, respectively.}
\label{fig:meander-designs}
\end{figure}

In order to verify the computed data in Fig.~\ref{fig:meander-designs-vs-qbound}, an antenna design sample with~$ka = 0.42$ was scaled to~$1.4$\,GHz and fabricated on a $25\,\mu$m-thick Polyimide film with $17\,\mu$m-thick copper foil bonded by an $25\,\mu$m-thick acrylic adhesive ($\varepsilon_\mathrm{r} = 4$). Thanks  to  the  very  thin  profile  of  the  substrate as  compared to wavelength, the effect of dielectric can be neglected for a radiation Q-factor evaluation. The input impedance of the prototype was measured using a differential technique~\cite{Meys_Diff_Method} and has been used to estimate the Q-factor via the~$Q_\mathrm{Z}$ formula~\cite{YaghjianBest_ImpedanceBandwidthAndQOfAntennas}. Radiation efficiency of the antenna was measured via a multiport near-field method~\cite{RazaYangHussain_MeasurementOfRadEff_2012} and was used to evaluate radiation Q-factor, and its confidence interval of width equal to two times the standard deviation, see triangular marker and corresponding error bar in Fig.~\ref{fig:meander-designs-vs-qbound}.

Figure~\ref{fig:meander-designs-vs-qbound} illustrates that a simple parametrization, such as the one from Fig.~\ref{fig:meander-parameterization}, is able to closely approach the radiation Q-factor bound limited to TM radiation~$Q_\mathrm{rad}^\mathrm{lb,TM}$ (see Box~\ref{B:Box1b}) in the entire frequency range of electrically small antennas. From this, it is possible to conclude that a complex design (\eg{}, the parameterizations found in \cite{Pantoja_2003_PreFractal}) is not needed to reach the lower bound.

The absolute lower bound for radiation Q-factor~$Q_\mathrm{rad}^\mathrm{lb}$ is unreachable by this \meanderline~antenna since its planar geometry and single feed scenario does not allow for an efficient excitation of combined TE and TM radiation. This contrasts to three-dimensional (\eg{}, spherical) geometries~\cite{Best2005,Thal2006}, where the dual mode behavior can be realized by a single feed network.

Parameters of the self-resonant designs from Fig.~\ref{fig:meander-designs-vs-qbound} are shown in Fig.~\ref{fig:meander-designs}. Design curves are fitted to the optimized parameters using a polynomial fit with good agreement. While some of the curves from Fig.~\ref{fig:meander-designs} can be found in~\cite{Fujimoto_Morishita_ModernSmallAntennas} for several parametrization, here all the designing curves are related back to Fig.~\ref{fig:meander-designs-vs-qbound} in which the Q-factor~$\Qrad$ is minimized. The presented data series can therefore be used for designing meandered dipoles approaching lower bounds on radiation Q-factor for TM antennas. It should, however, be noted that design curves from Fig.~\ref{fig:meander-designs} depends on the used parametrization and are valid only for~\mbox{$L/W = 2$} and~\mbox{$w/s \approx 1$}.

\subsection{Varying aspect ratios}
\label{sec:q-aspect-ratios}

Meander line antennas, introduced in the previous section, are now studied for various~$L/W$ and~$w/s$ aspect ratios and compared against the fundamental bounds calculated for each form factor.

\begin{figure}
\centering
\includegraphics[width=\figwidth]{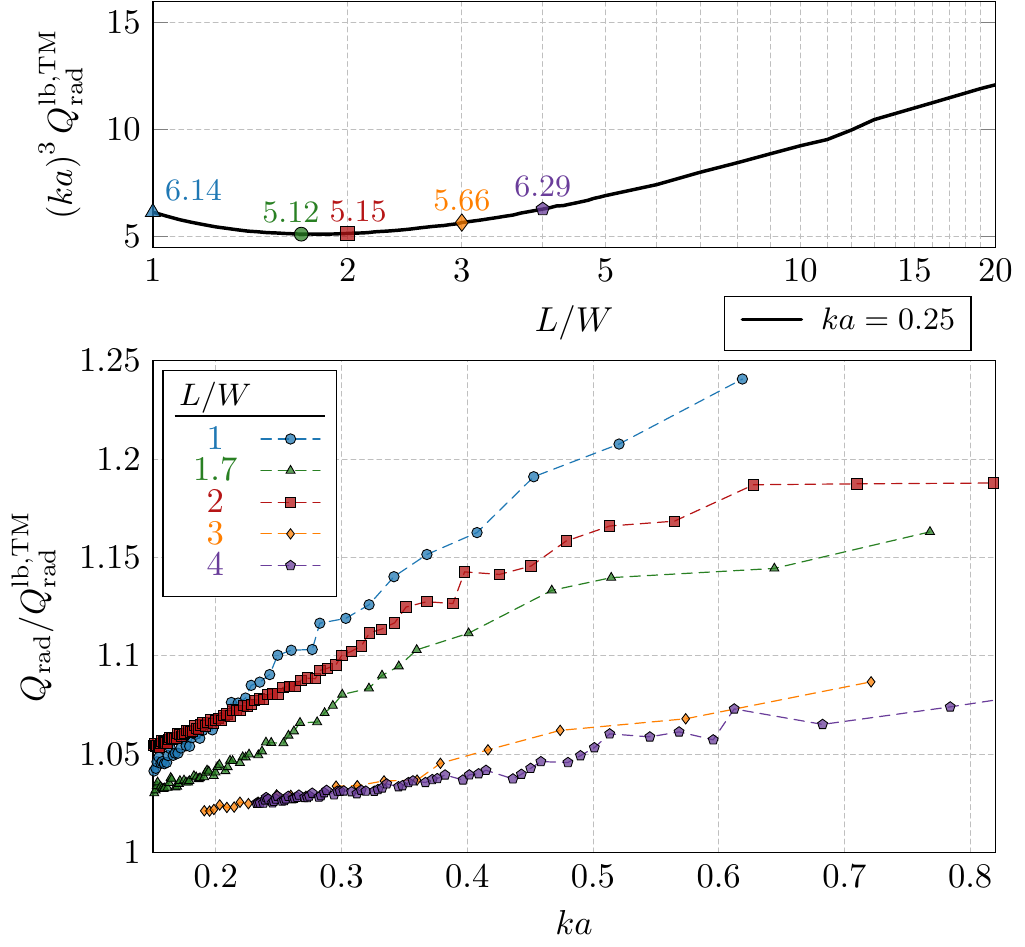}
\caption{Radiation Q-factor performance of PEC self-resonant meandered dipoles from Fig.~\ref{fig:meander-designs-vs-qbound} for various~$L/W$ aspect ratios. The radiation Q-factor is normalized to fundamental bound for a rectangular region which is shown in the top panel as a function of the aspect ratio. The minimum of the fundamental bound is found around ratio~\mbox{$L/W\approx 5/3$}. Generally, the higher the~$L/W$ ratio, the closer the \meanderline s are to the bound, however, at the cost of increasing absolute value of radiation Q-factor.}
\label{fig:meander-LWratios}
\end{figure}

In all cases, the value of radiation Q-factor~$\Qrad$ is normalized with respect to the minimal TM radiation Q-factor. Generally, Fig.~\ref{fig:meander-LWratios} shows that the minimal values can closely be approached for various~$L/W$ aspect ratios. Slightly better performance is observed for higher~$L/W$ ratios, however, at the cost of higher absolute bound on radiation Q-factor see top panel of Fig.~\ref{fig:meander-LWratios}.

\begin{figure}
\centering
\includegraphics[width=\figwidth]{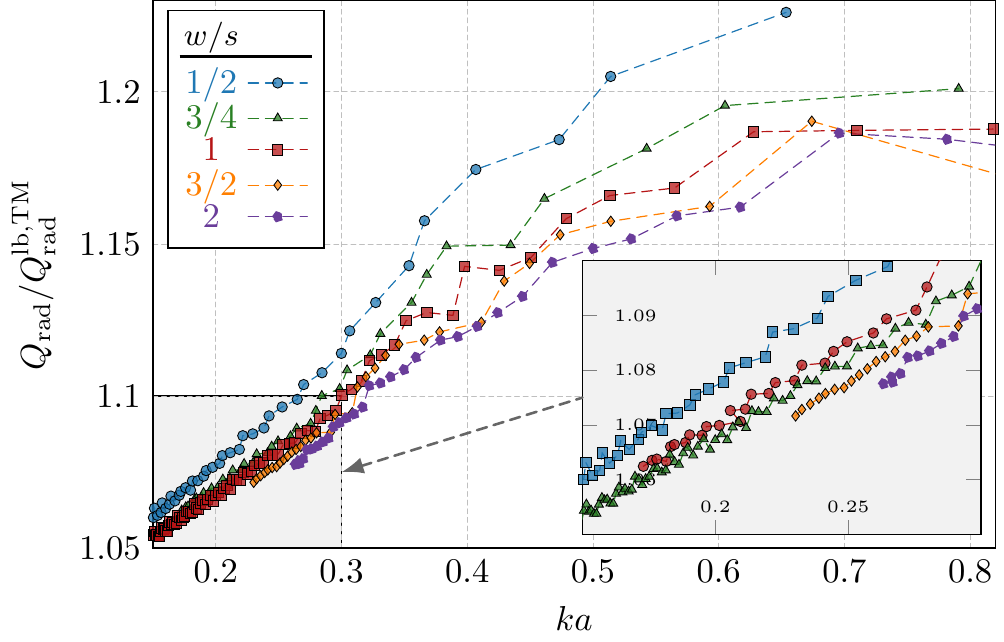}
\caption{The same study as in Fig.~\ref{fig:meander-LWratios} done for various ratios of \meanderline~width~$w$ to spacing~$s$.}
\label{fig:meander-wsatios}
\end{figure}

With respect to the varying~$w/s$ ratio, slightly better performance is observed for higher values, \ie{}, wider metallic strips. The differences become negligible for small values of~$ka$, see Fig.~\ref{fig:meander-wsatios}. Notice, however, that this behavior is substantially changed when ohmic losses are introduced, mainly since the spatial proximity of out-of-phased currents degrades the radiation and enhance the ohmic losses~\cite{BestMorrow_OnTheSignificanceOfCurrentVectorAlignmentInEstablishingTheResonantFrequencyOfSmallSpaceFillingWireAntennas}.

\subsection{The impact of impedance matching on Q-factor~$\Qrad$}

The designs obtained above are all self resonant ($\Xin \approx 0$), but no constraint was placed on the value of the input resistance~$\Rin$.  In most practical cases, the objective antenna input resistance is not driven by any antenna consideration but is set by the radio frequency electronic equipment to be interfaced with a particular antenna. Transmission lines and active receivers based on Low Noise Amplifiers (LNA) often require matching to~$50\,\Ohm$.  However, where devices with complex impedances are used, antenna resonance may not be ideal for conjugate matching and maximum power transfer. For example, a typical Power Amplifier (PA) output impedance is complex~\cite{Freescale1991}, with an input resistance lower than~$50\,\Ohm$ and an inductive (positive) reactive component. Similarly, passive RFID receivers based on Schottky diode rectifiers typically exhibit input resistances lower than~$50\,\Ohm$ and strong capacitive (negative) reactance~\cite{Marrocco_2008_ArtOfUHFRFID}.  Examples of nominal impedances~$\Zmatch$ for these systems are listed in Table~\ref{tab:impedances}.

\begin{table}
\centering
\caption{Three impedances of practical significance for antenna system design.}
\begin{tabular}{cc}
System & Input impedance $\Zmatch$ \\ \toprule
Power amplifier (PA) & $15+\J 50\Ohm$ \\
RFID chip (passive RX) & $20-\J 200\Ohm$ \\
Low-noise amplifier (LNA) & $50\Ohm$ \\ \bottomrule
\end{tabular}
\label{tab:impedances}
\end{table}

\begin{figure}
\includegraphics[width=\figwidth]{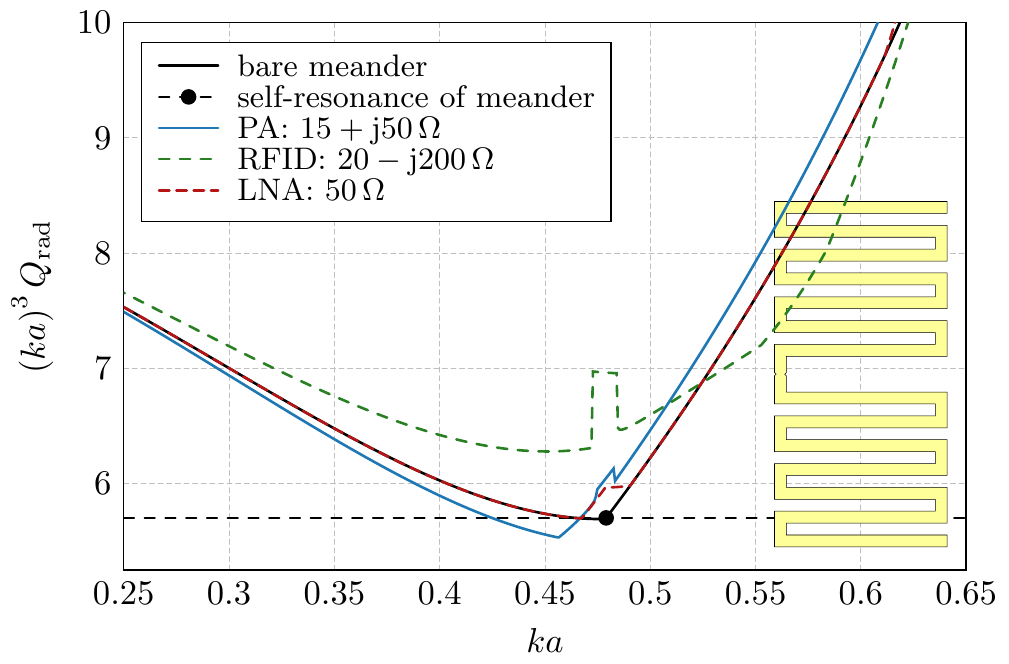}
\caption{Matched radiation Q-factor of a selected \meanderline~antenna for several impedance matching scenarios listed in Table~\ref{tab:impedances}.  Matching to each impedance is accomplished via a two-element reactive L-network, see inset in Fig.~\ref{fig:meander-complex-matching} for a schematic. In cases when several L-networks exist for a given matching impedance, the network with the lowest stored energy has been used. Abrupt jumps of the matched radiation \mbox{Q-factor} curves result from non-existence of matching by two inductances in certain frequency ranges. This double inductance matching is the most favorable scenario for a capacitive antenna.}
\label{fig:meander-matching-intro}
\end{figure}

\begin{figure}
\includegraphics[width=\figwidth]{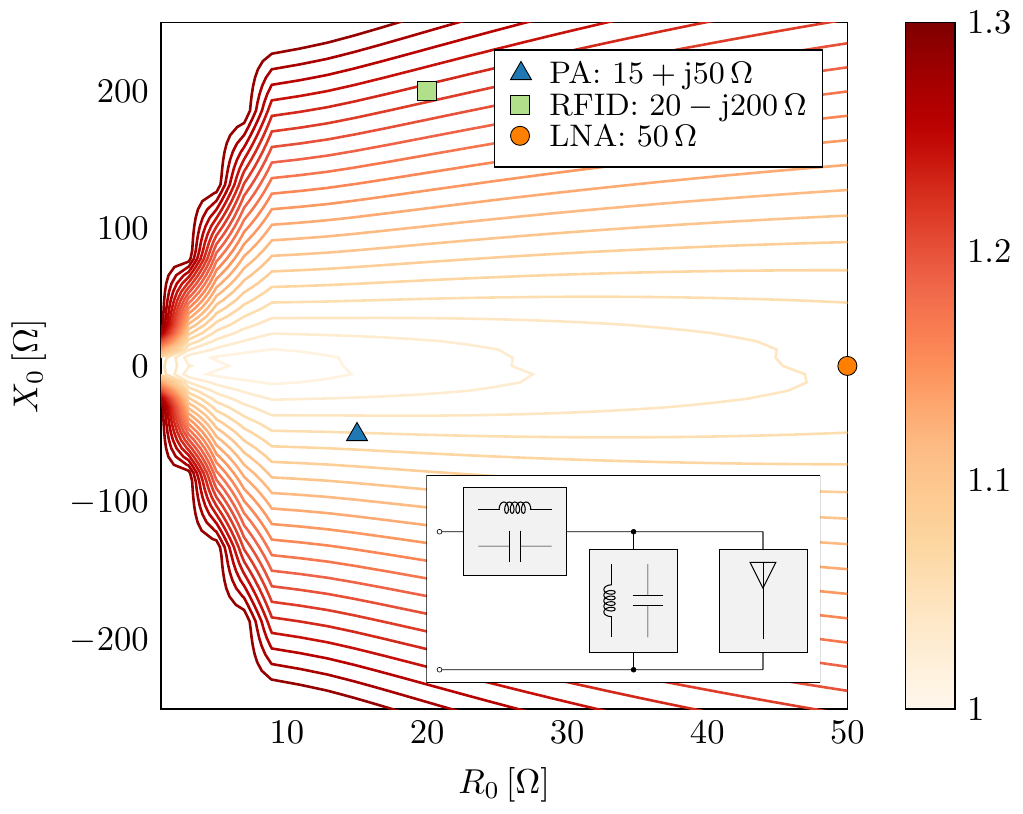}
\caption{Matched radiation Q-factor of a selected \meanderline~antenna for varying complex matching impedance normalized to radiation Q-factor of bare antenna. For each complex impedance, the \meanderline~is conjugate matched at its self-resonance (circle mark in Fig.~\ref{fig:meander-matching-intro}, $ka \approx 0.479$) using the lossless {L-network} matching circuit with lowest {Q-factor}. The markers denote the three impedances from Fig.~\ref{fig:meander-matching-intro} and Table~\ref{tab:impedances}.}
\label{fig:meander-complex-matching}
\end{figure}

Antennas may be designed to have input impedances which conjugate match a desired load. However, any of the designs shown in Fig.~\ref{fig:meander-designs-vs-qbound} can be conjugate matched to an arbitrary complex impedance~$\Zmatch$ through an L-network consisting of two reactive components~\cite{Pozar_MicrowaveEngineering}.  In many instances, the stored energies within these reactances will raise the radiation Q-factor of the system. To assess the cost of this form of simple matching, we select the design in Fig.~\ref{fig:meander-designs-vs-qbound} corresponding to self resonance at $ka = 0.479$. A set of lossless networks was generated to conjugate match the antenna to arbitrary complex impedances and the matched radiation Q-factors were calculated. A typical frequency dependence of this cost is depicted in Fig.~\ref{fig:meander-matching-intro} while the dependence on matching impedance is depicted in Fig.~\ref{fig:meander-complex-matching}. We observe that it is generally possible to transform the resonant antenna impedance to an arbitrary real value with minimal increase in radiation Q-factor, except when small resistance and high reactance is required.  As expected, adding a reactive component to the real-valued (resonant) antenna impedance necessarily increases radiation Q-factor, though this increase is on the order of $30\%$ for the most extreme of the three test impedances (RFID) examined here. Additionally, Fig.~\ref{fig:meander-matching-intro} shows that it is often possible to move slightly away from the self-resonant frequency and lower the overall radiation Q-factor by a small amount. Nonetheless, the minimum radiation Q-factor of the matched antenna is, for practical values of the matching impedance, within the vicinity of the self-resonance of the antenna.

\begin{figure}
\includegraphics[width=\figwidth]{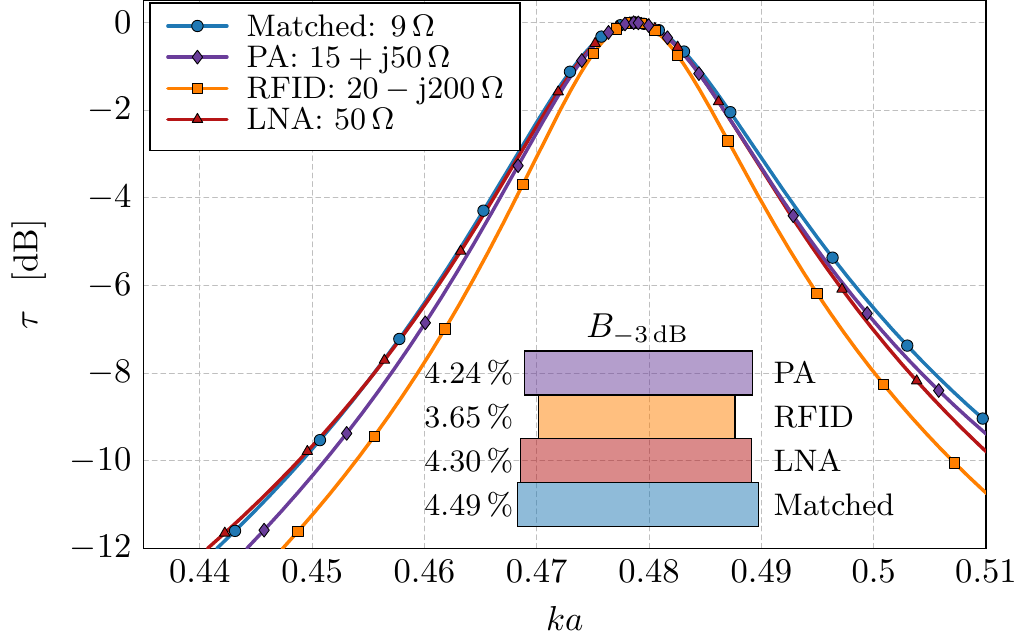}
\caption{Frequency dependence of normalized power delivered \mbox{$P_\mathrm{del}/P_\mathrm{cm}$} to the \meanderline~studied in Figs.~\ref{fig:meander-matching-intro} and~\ref{fig:meander-complex-matching}. The separate curves correspond to devices exhibiting the three practically relevant impedances from Table~\ref{tab:impedances} and to a device with impedance corresponding to that of the \meanderline~at its self-resonance (\Quot{Matched}).  In each case, the antenna is conjugate matched using an L-network at its self resonant frequency, $ka\approx0.479$.  The $-3~\mathrm{dB}$ bandwidths $50\%$ power delivered bandwidths $B_{-3~\mathrm{dB}}$ for each scenario are also listed.}
\label{fig:meander-realized-bandwidths}
\end{figure}

The importance of Q-factor is its relation to fractional bandwidth which is predicated on simple, single resonance behavior~\cite{YaghjianBest_ImpedanceBandwidthAndQOfAntennas}.  We demonstrate that the low variance in Q-factor corresponds to consistent realized bandwidth when L-networks are used to conjugate match an antenna to an arbitrary impedance.  Figure~\ref{fig:meander-realized-bandwidths} shows the power delivered $P_\mathrm{del}$ to the \meanderline~antenna studied above using a matched source ($\Zmatch = \Rin$) as well as with L-networks designed to match the antenna to the three complex impedances of practical interest in Table~\ref{tab:impedances}.  In each case, a network tunes the antenna to the desired (possibly complex) impedance at its natural resonant frequency. The frequency profile of the mismatch factor \cite{Pozar_MicrowaveEngineering,best2016optimizing}
\begin{equation}
\tau = \frac{P_\mathrm{del}}{P_\mathrm{cm}}=\frac{4R^\mathrm{m}_\mathrm{in}R_0}{|Z_\mathrm{in}^\mathrm{m}+\Zmatch|^2}
\end{equation}
is nearly identical in all four cases, in agreement with the predictions based on the relatively invariant Q-factor across these cases.  Here, $Z_\mathrm{in}^\mathrm{m}$ is the antenna impedance including the tuning network, $P_\mathrm{del}$ is the power delivered to the antenna, and $P_\mathrm{cm}$ is the power delivered under a conjugate match condition.  It is necessary to point out that we have assumed non-dispersive matching impedances, \ie{}, $\partial \Zmatch / \partial \omega = 0$.  In practice, the matching impedance may be dispersive within the band of interest, in which case the relation between Q-factor and bandwidth described in \cite{YaghjianBest_ImpedanceBandwidthAndQOfAntennas} ceases to be valid.  However, inclusion of a dispersive load impedance may not necessarily cause major changes to the realized bandwidth due to the already heavily frequency-dependent nature of the impedance of high Q-factor antennas. Despite this simplification, when generating Fig.~\ref{fig:meander-realized-bandwidths}, lumped inductors and capacitors in each tuning network are modeled as frequency dependent impedances.

The results in Figs.~\ref{fig:meander-complex-matching} and~\ref{fig:meander-realized-bandwidths} numerically suggest that there is little cost in bandwidth to match a self-resonant antenna to arbitrary impedances. However, further considerations reveal why it is of practical importance to design an antenna with a given impedance, rather than relying on this form of matching.  First, the use of lumped components increases complexity and cost of an antenna system and the required component values for the L-networks described in this section may not be realizable.  Second, lumped components made of any practical, lossy material (\eg{}, metallic inductors) increase the net loss in an antenna system while not adding any potential radiation mechanism.  This guarantees a decrease in overall efficiency, particularly in high Q-factor antennas~\cite{Smith_1977_TAP}.  Additionally, tunability or the use of broadband multiple resonance matching may benefit from the design of an antenna with specific impedance characteristics, \eg{}, to increase the radiation resistance \cite{best2008small,Best_ElectricallySmallResonantPlanarAntennas}.

\section{Radiation efficiency of Q-optimal antennas}
\label{sec:Efficiency-optimization}

The previous section demonstrated that \meanderline~antennas are nearly optimal with respect to radiation Q-factor, including the cases when matching to realistic complex impedances is desired. This section studies how these antennas perform with respect to another critical antenna metric: radiation efficiency (see Box~\ref{B:delta1}).  Specifically, we examine their performance with respect to radiation efficiency bounds (see Box~\ref{B:delta2}).

\begin{figure}
\includegraphics[width=\figwidth]{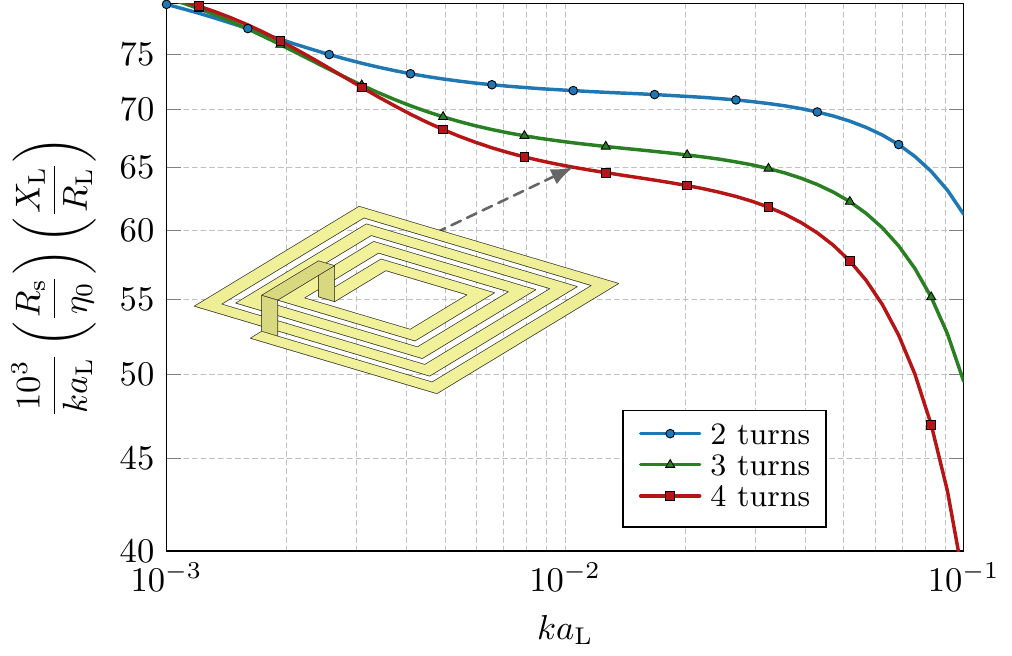}
\caption{Normalized reactances of selected rectangular spiral inductors. The quantities $\XL$, $\RL$, and $\Rsurf$ denote input reactance, input resistance, and surface resistance, respectively. The radius $a_\mathrm{L}$ defines the smallest sphere circumscribing the inductor.}
\label{fig:QL}
\end{figure}

Before presenting the radiation efficiency of matched \meanderline~antennas it is necessary to deal with losses in the matching circuit since, similarly to the case of Q-factor, any matching circuit with finite losses will worsen the overall efficiency of the antenna system. Throughout this section we will assume that all matching networks are composed of lossless capacitors and lossy inductors\footnote{Q-factors of lossy capacitors are typically much higher than those of lossy inductors.}. The inductors are further assumed to be planar, made of the same material (metallic sheet, surface resistivity~$\Rsurf$) as the antenna itself. Under such restrictions it is possible to estimate the loss added by a matching network quite precisely using data from Fig.~\ref{fig:QL}, which shows the normalized reactance, $\left(10^3/ka_L\right)\left(\Rsurf/\ZVAC \right) \left(\XL / \RL \right)$, of several spiral inductors as a function of their electrical size.  Here $\ZVAC$ denotes the free space impedance. The normalized reactance in Fig.~\ref{fig:QL} is independent on surface resistance $R_s$ and, at small electrical size, just weakly dependent on number of turns and frequency, consistent with classical relations for helical air-core inductors~\cite{nagaoka1909inductance}. A conservative value $\left(10^3/ka_\mathrm{L}\right)\left(\Rsurf/\ZVAC \right) \left(\XL / \RL  \right) = 66$ will be used in this section to determine losses of all inductors within the L-matching network, assuming further that inductors are always ten times smaller in electrical size than the antenna, \ie{}, \mbox{$a_L=a/10$}. This last assumption enforces the use of an electrically small, approximately lumped element, matching network.

\begin{figure}
\includegraphics[width=\figwidth]{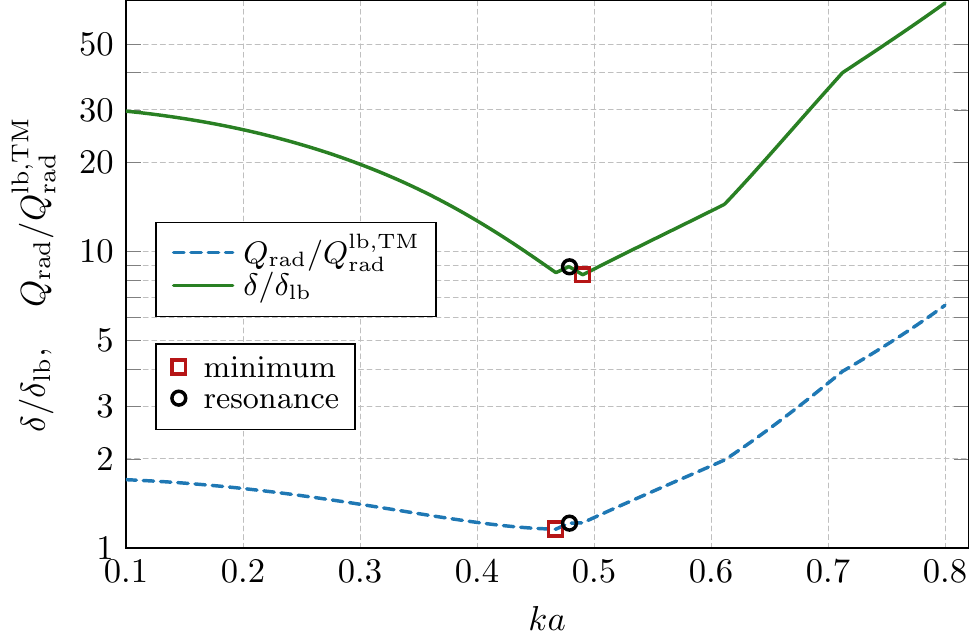}
\caption{Normalized radiation Q-factor and normalized dissipation factor of a selected \meanderline~antenna matched to~$50\Ohm$ over a range of frequencies (scaled here as electrical size~$ka$). The self-resonance of the antenna (\mbox{$ka=0.47$}) is denoted on each trace with a circular marker while the minimum of each trace is also marked. The same data are also plotted as a curve parameterized by frequency in Fig.~\ref{fig:efficiency_explanation_2}.}
\label{fig:efficiency_explanation_1}
\end{figure}

Lossy elements with the above mentioned specifications are used to match the meander studied in Figs.~\ref{fig:meander-matching-intro}--\ref{fig:meander-realized-bandwidths} to impedance $\Zmatch = 50\Ohm$ over a band of interest near the \meanderline 's self-resonant frequency.  The resulting radiation Q-factor and efficiency (here presented in the form of dissipation factor,~$\delta$) are depicted in Fig.~\ref{fig:efficiency_explanation_1} as functions of frequency (scaled as electrical size~$ka$). The figure reiterates the previously-observed near-optimal performance of \meanderline~antennas with respect to radiation Q-factor, but, surprisingly, shows a rather poor performance with respect to radiation efficiency. This metric is, at the self-resonance frequency of the antenna, almost one order of magnitude worse than the value of the physical bound (see Box~\ref{B:delta2}). Similarly to radiation Q-factor, dissipation factor reaches its minimum in the vicinity of the resonance frequency, at least in the case of realistic values of matching impedances used here.

\begin{figure}
\includegraphics[width=\figwidth]{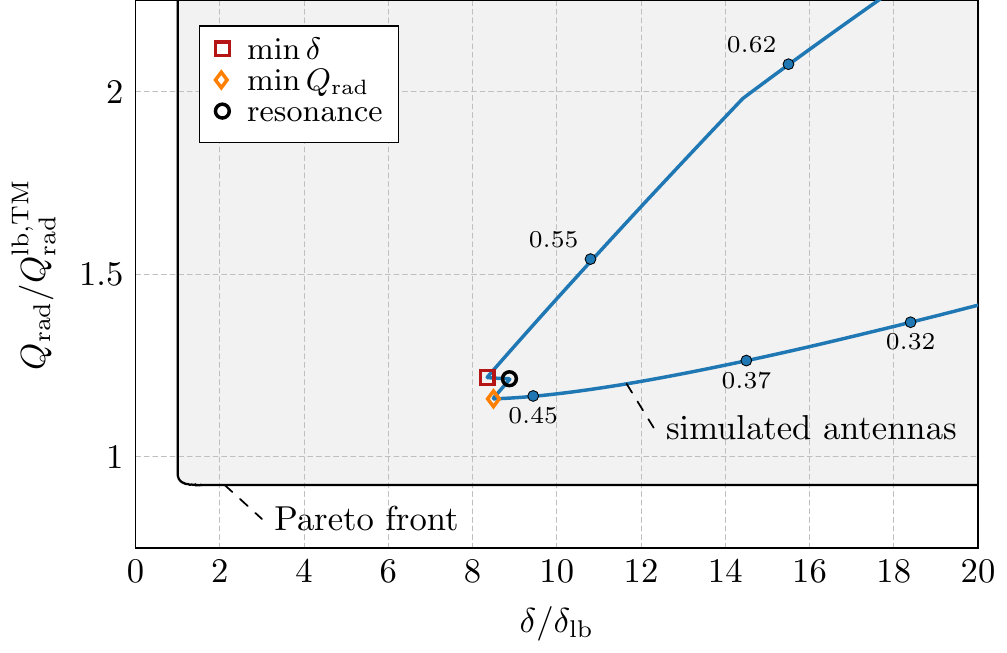}
\caption{A two-dimensional view of the lower bounds on dissipation factor and radiation Q-factor. A black solid curve denotes the Pareto front describing the trade-off between these quantities and the corresponding feasible region (shaded). The data from Fig.~\ref{fig:efficiency_explanation_1} are drawn as a curve parameterized by frequency, with the relevant points in Fig.~\ref{fig:efficiency_explanation_1} being similarly marked. The circle markers along the curve representing the simulated antenna show the electrical size $ka$.}
\label{fig:efficiency_explanation_2}
\end{figure}

Within the used normalization of dissipation factor and radiation Q-factor, it is reasonable to represent the data from Fig.~\ref{fig:efficiency_explanation_1} as a two dimensional curve (radiation Q-factor vs. dissipation factor) parametrized by frequency, see Fig.~\ref{fig:efficiency_explanation_2}. The figure also shows the Pareto front (represented by the black line) evaluated by the method from \cite{GustafssonCapekSchab_Tradeoffs2017}, which demonstrates the optimal trade-off between radiation Q-factor and dissipation factor for the given design geometry and frequency. The Pareto front has been evaluated at \mbox{$ka=0.5$}, but, due to the used normalization, it is almost independent of electrical size. The Pareto front was evaluated for a combination of TM and TE modes which, as normalized to the TM bound $\QradminTM$, gives values lower than one. The reason for this particular normalization is that TM bounds represent meaningful limit of one-port planar antennas.

\begin{figure}
\includegraphics[width=\figwidth]{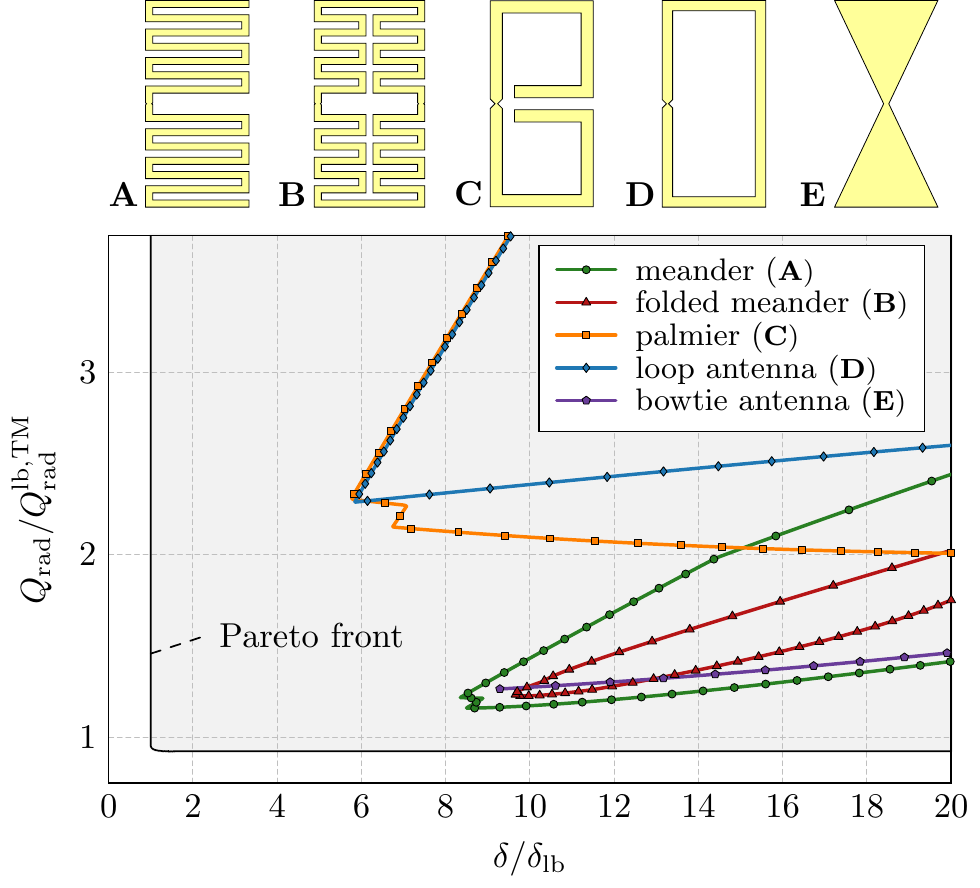}
\caption{Two dimensional frequency parametrized plot representing the physical bound and several antennas. The matching impedance equals to~$50\Ohm$. The tip of each curve lies in the vicinity of resonance or anti-resonance of the antenna. The electrical sizes at the resonance or anti-resonance frequency are: A ($ka=0.48$, res.), B ($ka=0.85$, res.), C ($ka=0.50$, res.), D ($ka=0.59$, anti-res.), E (no resonance).}
\label{fig:pareto-trade-offs}
\end{figure}

The two-dimensional plot in Fig.~\ref{fig:efficiency_explanation_2} represents a complete comparison of various antenna designs with respect to matched efficiency and matched radiation {Q-factor}. An example of such comparison is shown in Fig.~\ref{fig:pareto-trade-offs}, where the normalized and frequency-parameterized $Q$--$\delta$ curves are drawn for several small antenna designs within the same design specifications\footnote{The bounding geometry, material parameters, and restrictions on matching network topology and losses are all kept constant across each design.}. Figure~\ref{fig:pareto-trade-offs} clearly presents the superior performance in efficiency and {Q-factor} of simple \meanderline~antennas shown in Fig.~\ref{fig:meander-parameterization} with respect to other designs. It also shows that although there exist other \meanderline s which perform slightly better in radiation efficiency (Palmier pastry type,~\cite{Palmier}) this improvement costs much in the radiation {Q-factor}. In conclusion, simple \meanderline~antennas present the best trade-off between radiation Q-factor and dissipation factor from the depicted antennas when matching to real impedances is demanded.  As in the previous section, we note that the use of more advanced matching topology (\eg{}, folding or impedance transformer) may benefit from alternative antenna designs.

Figures~\ref{fig:efficiency_explanation_2}~and~\ref{fig:pareto-trade-offs} show that the considered antenna structures, which are close to optimal in radiation {Q-factor}, are far away from the efficiency bounds. This is puzzling since resonant modes optimal in radiation {Q-factor} and efficiency are similar in nature.  However, there are important differences. Radiation {Q-factor} restricted to TM modes is minimized by separation of charges and inducing dipole like currents~\cite{GustafssonTayliEhrenborgEtAl_AntennaCurrentOptimizationUsingMatlabAndCVX}. These modes can be tuned to resonance by inducing edge loops along the structure. TM efficiency, on the other hand, is minimized by inducing homogeneous currents \cite{2018_Shahpari_TAP}. These are similar in nature to the dipole like currents minimizing {Q-factor}, but the loop currents which minimize TE {Q-factor} and maximize TE efficiency are fundamentally different. Where low {Q-factor} loops tend to be confined towards the edges of the structure, high efficiency loops are spread across the whole area~\cite[Fig.~4]{GustafssonCapekSchab_Tradeoffs2017}. Such loop currents are naturally restricted as an original simply connected object fully filling a prescribed bounding box is perforated, forcing the current distribution into more inhomogeneous forms. Thus, low {Q-factor} loops are tolerant of alterations to a structure whereas high efficiency loops are harshly disrupted.

\begin{figure}
\includegraphics[width=1\linewidth]{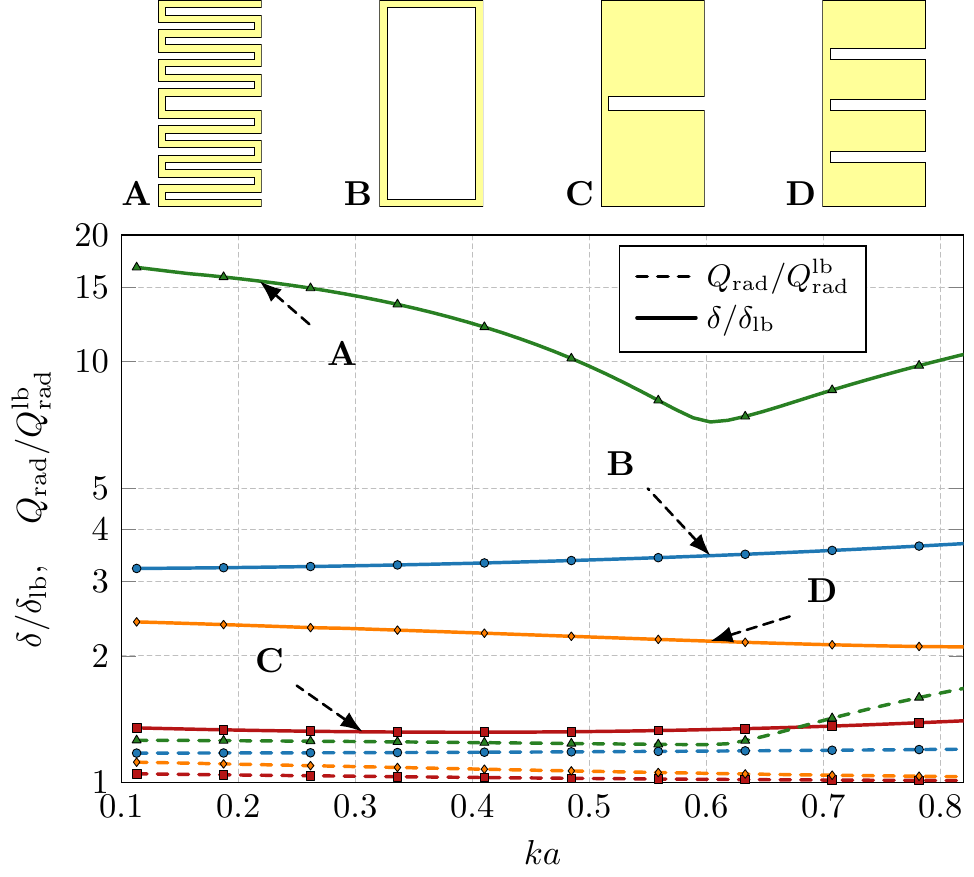}
\caption{Illustration of how much different metrics deteriorate when a solid plate is permuted. The figure shows four generic shapes, a single slot, three slots, a \meanderline, and a loop. The ratio between optimal loss factor for different geometries and the optimal loss factor for the full plate are shown in solid lines. The ratio between the optimal mixed mode Q-factor for the same geometries and the optimal mixed mode Q-factor of the full plate are shown in dashed lines.}
\label{fig:cut_metal}
\end{figure}

In Fig.~\ref{fig:cut_metal}, the optimal resonant {Q-factor} and dissipation factor are plotted normalized to the corresponding bounds of a rectangular plate. Data for different shapes made by removing portions of the plate are shown. The currents on the structures in Fig.~\ref{fig:cut_metal} have been calculated with current optimization without physical feeding. It is clear that removing metal does not greatly affect the achievable radiation {Q-factor}, at worst reducing it to the TM-only bound. However, when metal is removed from the plate the loss factor is significantly increased, especially for small electrical sizes. Thus, while optimal radiation {Q-factor} and radiation efficiency modes are fairly similar, removing design space has a much greater effect on the loss factor than the {Q-factor} in relation to the physical bounds. This can be seen in Fig.~\ref{fig:cut_metal} where the loss factor of the optimal resonant currents is very high for the structures with slots in them.  Consider the \meanderline~antenna which has significantly higher loss factor at electrical sizes~\mbox{$ka<0.4$}, here the loop modes are extremely disrupted, however, the Q-factor is hardly affected. The sharp change in the \meanderline 's loss at around~\mbox{$ka=0.6$} is due to its resonance, where it is possible to induce a resonant dipole mode on the structure.  This example illustrates a fundamental challenge in designing efficient small resonant antennas: many of the strategies normally utilized to induce resonance, such as meandering, harshly limit the achievable efficiency. 

\begin{boxexampl}[label={B:delta1}]{Radiation efficiency}
The radiation efficiency of an antenna is defined as
\begin{equation}
\label{Box_eta}
\eta = \frac{\Prad}{\Prad + \Plost} = \frac{1}{1 + \DissipF},
\end{equation}
where, as in~\eqref{eq:qdef_box1a}, $\Prad$ and $\Plost$ are radiated and ohmic dissipated power, respectively, and \mbox{$\delta = \Plost/ \Prad$} is the dissipation factor~\cite{Harrington_EffectsOfAntennaSizeOnGainBWandEfficiency}. Along with bandwidth radiation efficiency is a key antenna performance parameter, particularly in electrically small systems where it is known to decrease rapidly with antenna size.\\

For objects with homogeneous loss properties, \eg{}, uniform surface resistance or conductivity, the dissipation factor~$\delta$ is a linear function of those properties.  As such, values of dissipation factor can be normalized by surface resistivity for ease of comparison.   
\end{boxexampl}

\begin{boxexampl}[label={B:delta2}]{Lower bounds to dissipation factor}
Two different paradigms for minimization of dissipation factor exist. The first assumes that tuning or general impedance matching of the antenna can be performed in a lossless manner. Under this assumption, the optimal current density minimizing dissipation factor is the result of a generalized eigenvalue problem \cite{UzsokySolymar_TheoryOfSuperDirectiveLinearArrays,Harrington_EffectsOfAntennaSizeOnGainBWandEfficiency,Harrington_AntennaExcitationForMaximumGain,2018_Shahpari_TAP}. Such lower bounds were shown to scale with electrical size as $(ka)^{-2}$ and are straightforward to calculate. Their major drawback, however, is that, by neglecting matching network losses, the resulting dissipation factors are overly optimistic and unachievable by realistic designs where some form of matching is required \cite{Smith_1977_TAP}.\\

One solution to the aforementioned drawback is a paradigm in which the optimal currents are calculated while taking into account the dissipation cost of achieving resonance or general matching \cite{Smith_1977_TAP, Jelinek2018TAPCost}. Dissipation factors coming from this second paradigm are generally closer to realistic designs and scale with electrical size as~$(ka)^{-4}$ \cite{Jelinek2018TAPCost,Pfeiffer_FundamentalEfficiencyLimtisForESA,Thal2018,GustafssonCapekSchab_Tradeoffs2017}. Lower bounds to tuned dissipation factor for several selected shapes are shown in Table~\ref{Tab:Bounds}.
\end{boxexampl}

\section{Antennas optimal in other parameters}

\begin{figure}[h]
\centering
\includegraphics[width=\figwidth]{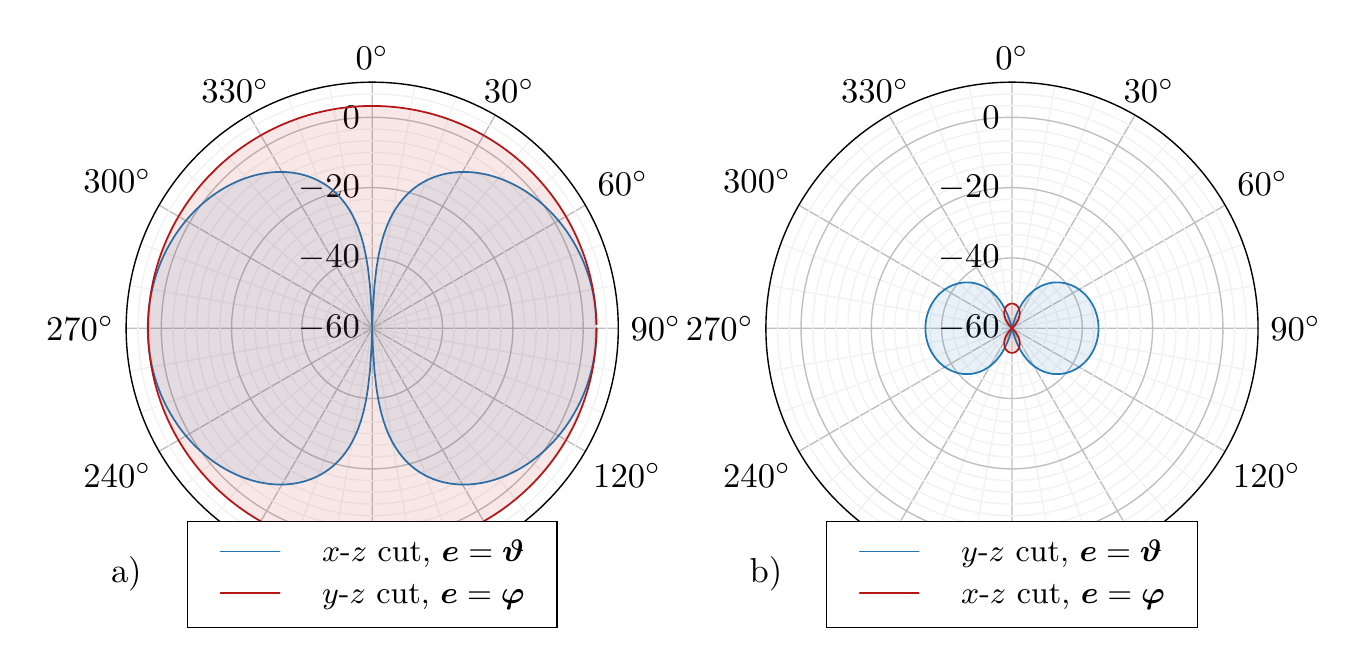}
\caption{Directivity (in dB scale) with respect to an isotropic radiator of a meander line antenna depicted in Fig.~\ref{fig:meander-designs-vs-qbound} as the very right inset. Antenna is placed in $x$-$z$ plane, with a longer side aligned with $z$-direction.}
\label{fig:radPatMeanderline}
\end{figure}

Determining the best possible Q-factor can be formulated as a minimization problem. Therefore it is possible to add different or additional constraints to such an optimization. So far, in this paper, we have considered the constraints of efficiency and impedance matching. Another type of constraints are different kinds of field-shaping requirements of near and/or far-fields~\cite{Gustafsson+Nordebo2013,Jonsson+etal2017b}. For small antennas it is well known that the radiated far-field tends to resemble a dipole pattern, meander line antenna treated in this paper being no exception, see Fig.~\ref{fig:radPatMeanderline}. However, with these types of Q-factor optimization procedures it is possible to determine the Q-factor cost, to have the antenna radiating with a certain front-to-back ratio or (super-) directivity in a given direction. These classes of bounds indicate that for a limited bandwidth cost it is possible to extend, \eg{}, the directivity beyond the traditional dipole pattern, see ~\cite{Gustafsson+Nordebo2013,Jonsson+etal2017b,Ferrero+etal2017,Shi+etal2017,Yaghjian+etal2008,Pigeon+etal2014,Ziolkowski+etal2013,Kim+etal2012}.

\begin{figure}
\includegraphics[width=\figwidth]{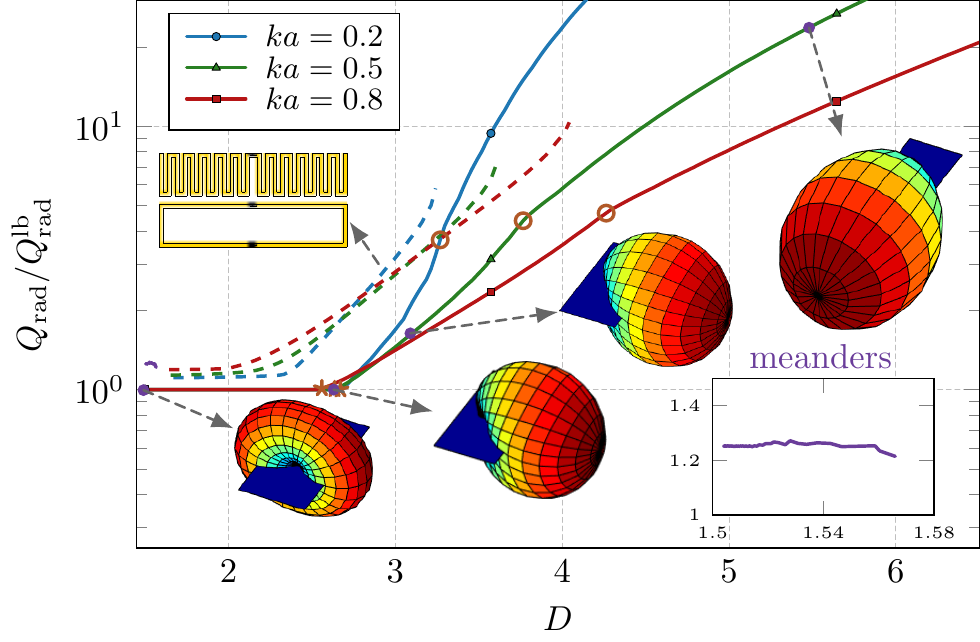}
\caption{The cost in Q-factor for a desired directivity in a lossless 2:1 shaped antenna. The three line colors represent electrical sizes \mbox{$ka=\left\{0.2,0.5,0.8\right\}$} and the radiation patterns correspond to the $ka = 0.5$ case for the optimal current. All $Q$- vs~$D$-values for the \meanderline s of Fig.~\ref{fig:meander-designs-vs-qbound} are represented in purple, and the relevant region is zoomed in the right bottom inset. Dashed curves show the results for an array composed of a \meanderline~antenna element with one feed and a loop antenna element with two feeds.}
\label{fig:QvsD}
\end{figure}

To illustrate bounds on superdirectivity, Q-factor optimization for a given directivity described in~\cite{Gustafsson+Nordebo2013,Jonsson+etal2017b}, was solved for a small antenna with length to width ratio of 2:1, infinitesimal thickness, and electrical sizes \mbox{$ka\in\left\{0.2,0.5,0.8\right\}$}. 
The bounds for low directivities are identical to the lower bound on the Q-factor, where the radiation pattern changes from that of an elliptically polarized dipole with $D\approx 1.5$ to that of a Huygens source with directivity just below $D=3$ and the main beam pointing in the direction of the longest side~\cite{CapekGustafssonSchab_MinimizationOfAntennaQualityFactor}. Higher directivities require quadrupole and higher order modes which increases the Q-factor  rapidly~\cite{Chu_PhysicalLimitationsOfOmniDirectAntennas}. The direction of the main beam changes from the longest side of the antenna to an endfire pattern along the shortest side at $\Qrad/\QradminTM\approx 3$ as indicated by circles in Fig.~\ref{fig:QvsD}.

Much like bounds on other parameters (\eg{}, efficiency), it is an open problem if the directivity-constrained limits are reachable for all sizes and desired directivity, even under idealized lossless conditions. As a demonstration of one possible high directivity, low Q-factor design, a three port array composed of a \meanderline~and a loop structure with optimized feeding is presented in relation to the bounds, see Fig.~\ref{fig:QvsD}. 
However, high directivity for single port antennas remains, as of yet, far from the bound and new designs ideas that allow a high directivity with larger bandwidth are desired.   

\section{Conclusion}
The possibilities how to approach the fundamental bounds on selected antenna metrics were investigated. A planar region of rectangular form factor was considered. It was observed that the lower bound on Q-factor with radiation restricted to TM modes only is closely approached by a \meanderline{}~antenna for a broad range of electrical sizes. The optimal design parameters were depicted and various aspect ratios of the bounding rectangle were studied together with selected ratios of the strip and slot widths. The simulated results were verified by a measurement of a fabricated prototype. The impedance matching and its impact on the Q-factor of the antenna was studied, concluding that the effect of the impedance matching on radiation Q-factor is minor and, in some cases, that matching the antenna slightly away from its self-resonance can even decrease its Q-factor. Radiation efficiency of the \meanderline{}~antennas optimal in Q-factor was evaluated, taking into account ohmic losses dissipated in the matching circuit. It was observed that the radiation efficiency of the studied \meanderline{}~antennas is far from an upper bound of a rectangular patch. Several other planar antennas were similarly evaluated against fundamental bounds yielding consistent conclusions: synthesizing antenna designs which approach the upper bound on radiation efficiency is more difficult than designing those which reach the lower bound on Q-factor. The reason was identified in the high sensitivity of radiation efficiency to the perturbation of ideal constant current density. Namely, when an initial structure fully filling the prescribed bounding box is perforated (as is done in a practical synthesis procedure), the performance of maximum efficiency current distributions drops much faster than that of a minimum \mbox{Q-factor} distribution. Finally, a Pareto-type bound between \mbox{Q-factor} and directivity has been calculated and compared to \meanderline{}~antennas. An attempt has been made to find an antenna with reasonably low Q-factor and directivity higher than that of an electric dipole type antenna. Nevertheless, no planar antenna with one feed fulfilling these contradictory constraints was found. This task and its feasibility remains as a subject for ongoing research.

The fundamental bounds, \ie{}, the lower bounds on Q-factor, the upper bounds on radiation efficiency, the Pareto-optimality between Q-factor and efficiency, or Q-factor and directivity, were demonstrated to be powerful tools for judging the performance of the radiating devices. If the realistic designs are compared to the fundamental bounds, designer can assess how far from the optima the design is, therefore, if further improvement is needed. Furthermore, incremental progress in design improvement can be put into context by considering the remaining distance between an antenna's realized performance and the fundamental bounds. It is the normalized ratio of the actual device's performance to the fundamental bounds what reveals the real quality of the design.

\bibliographystyle{IEEEtran}
\bibliography{references}

\begin{thebibliography}{10}
\providecommand{\url}[1]{#1}
\csname url@samestyle\endcsname
\providecommand{\newblock}{\relax}
\providecommand{\bibinfo}[2]{#2}
\providecommand{\BIBentrySTDinterwordspacing}{\spaceskip=0pt\relax}
\providecommand{\BIBentryALTinterwordstretchfactor}{4}
\providecommand{\BIBentryALTinterwordspacing}{\spaceskip=\fontdimen2\font plus
\BIBentryALTinterwordstretchfactor\fontdimen3\font minus
  \fontdimen4\font\relax}
\providecommand{\BIBforeignlanguage}[2]{{%
\expandafter\ifx\csname l@#1\endcsname\relax
\typeout{** WARNING: IEEEtran.bst: No hyphenation pattern has been}%
\typeout{** loaded for the language `#1'. Using the pattern for}%
\typeout{** the default language instead.}%
\else
\language=\csname l@#1\endcsname
\fi
#2}}
\providecommand{\BIBdecl}{\relax}
\BIBdecl

\bibitem{RahmatMichielssen_ElectromagneticOptimizationByGenetirAlgorithms}
Y.~Rahmat-Samii and E.~Michielssen, Eds., \emph{Electromagnetic Optimization by
  Genetic Algorithm}.\hskip 1em plus 0.5em minus 0.4em\relax Wiley, 1999.

\bibitem{Haupt+Werner2007}
R.~L. Haupt and D.~H. Werner, \emph{Genetic Algorithms in
  Electromagnetics}.\hskip 1em plus 0.5em minus 0.4em\relax Wiley-IEEE Press,
  2007.

\bibitem{RahmatSamii_Kovitz_Rajagopalan-NatureInspiredOptimizationTechniques}
Y.~Rahmat-Samii, J.~M. Kovitz, and H.~Rajagopalan, ``Nature-inspired
  optimization techniques in communication antenna design,'' \emph{Proc. IEEE},
  vol. 100, no.~7, pp. 2132--2144, July 2012.

\bibitem{OnwuboluBabu_NewOptimizationTechniquesInEngineering}
G.~C. Onwubolu and B.~V. Babu, \emph{New Optimization Techniques in
  Engineering}.\hskip 1em plus 0.5em minus 0.4em\relax Springer, 2004.

\bibitem{Deb_MultiOOusingEA}
K.~Deb, \emph{Multi-Objective Optimization using Evolutionary
  Algorithms}.\hskip 1em plus 0.5em minus 0.4em\relax Wiley, 2001.

\bibitem{CismasuGustafsson_FBWbySimpleFreuqSimulation}
M.~Cismasu and M.~Gustafsson, ``Antenna bandwidth optimization with single
  frequency simulation,'' \emph{IEEE Trans. Antennas Propag.}, vol.~62, no.~3,
  pp. 1304--1311, 2014.

\bibitem{GustafssonSohlKristensson_IllustrationsOfNewPhysicalBoundOnLinearlyPolAntennas}
M.~Gustafsson, C.~Sohl, and G.~Kristensson, ``Illustrations of new physical
  bounds on linearly polarized antennas,'' \emph{IEEE Trans. Antennas Propag.},
  vol.~57, no.~5, pp. 1319--1327, May 2009.

\bibitem{Best_ElectricallySmallResonantPlanarAntennas}
S.~R. Best, ``Electrically small resonant planar antennas,'' \emph{IEEE
  Antennas Propag. Mag.}, vol.~57, no.~3, pp. 38--47, June 2015.

\bibitem{IEEEStd_antennas}
\emph{145-2013 -- IEEE Standard for Definitions of Terms for Antennas}, IEEE
  Std., March 2014.

\bibitem{Chu_PhysicalLimitationsOfOmniDirectAntennas}
L.~J. Chu, ``Physical limitations of omni-directional antennas,'' \emph{J.
  Appl. Phys.}, vol.~19, pp. 1163--1175, 1948.

\bibitem{Fante1969}
R.~L. Fante, ``Quality factor of general antennas,'' \emph{IEEE Trans. Antennas
  Propag.}, vol.~17, no.~2, pp. 151--155, Mar. 1969.

\bibitem{YaghjianBest_ImpedanceBandwidthAndQOfAntennas}
A.~D. Yaghjian and S.~R. Best, ``Impedance, bandwidth and {Q} of antennas,''
  \emph{IEEE Trans. Antennas Propag.}, vol.~53, no.~4, pp. 1298--1324, April
  2005.

\bibitem{Collin+Rothschild1964}
R.~E. Collin and S.~Rothschild, ``Evaluation of antenna {Q},'' \emph{IEEE
  Trans. Antennas Propag.}, vol.~12, pp. 23--27, Jan. 1964.

\bibitem{Harrington+Mautz1972}
R.~F. Harrington and J.~R. Mautz, ``Control of radar scattering by reactive
  loading,'' \emph{IEEE Trans. Antennas Propag.}, vol.~20, no.~4, pp. 446--454,
  1972.

\bibitem{Rhodes1976}
D.~R. Rhodes, ``Observable stored energies of electromagnetic systems,''
  \emph{Journal of the Franklin Institute}, vol. 302, no.~3, pp. 225--237,
  1976.

\bibitem{Collin1998}
R.~E. Collin, ``Minimum {Q} of small antennas,'' \emph{J. Electromagnet. Waves
  Appl.}, vol.~12, pp. 1369--1393, 1998.

\bibitem{Vandenbosch_ReactiveEnergiesImpedanceAndQFactorOfRadiatingStructures}
G.~A.~E. Vandenbosch, ``Reactive energies, impedance, and {Q} factor of
  radiating structures,'' \emph{IEEE Trans. Antennas Propag.}, vol.~58, no.~4,
  pp. 1112--1127, Apr. 2010.

\bibitem{GustafssonCismasuJonsson_PhysicalBoundsAndOptimalCurrentsOnAntennas_TAP}
M.~Gustafsson, M.~Cismasu, and B.~L.~G. Jonsson, ``Physical bounds and optimal
  currents on antennas,'' \emph{IEEE Trans. Antennas Propag.}, vol.~60, no.~6,
  pp. 2672--2681, June 2012.

\bibitem{Geyi2011}
W.~Geyi, \emph{Foundations of Applied Electrodynamics}.\hskip 1em plus 0.5em
  minus 0.4em\relax John Wiley \& Sons, 2011.

\bibitem{Gustafsson+Jonsson2015b}
M.~Gustafsson and B.~L.~G. Jonsson, ``Stored electromagnetic energy and antenna
  {Q},'' \emph{Progress In Electromagnetics Research (PIER)}, vol. 150, pp.
  13--27, 2015.

\bibitem{CapekGustafssonSchab_MinimizationOfAntennaQualityFactor}
M.~Capek, M.~Gustafsson, and K.~Schab, ``Minimization of antenna quality
  factor,'' \emph{IEEE Transactions on Antennas and Propagation}, vol.~65,
  no.~8, pp. 4115--4123, 2017.

\bibitem{Gustafsson+Jonsson2015a}
M.~Gustafsson and B.~L.~G. Jonsson, ``Antenna {Q} and stored energy expressed
  in the fields, currents, and input impedance,'' \emph{IEEE Trans. Antennas
  Propag.}, vol.~63, no.~1, pp. 240--249, 2015.

\bibitem{Schab+etal2018}
K.~Schab, L.~Jelinek, M.~Capek, C.~Ehrenborg, D.~Tayli, G.~A. Vandenbosch, and
  M.~Gustafsson, ``Energy stored by radiating systems,'' \emph{IEEE Access},
  vol.~6, pp. 10\,553 -- 10\,568, 2018.

\bibitem{VolakisChenFujimoto_SmallAntennas}
J.~L. Volakis, C.~Chen, and K.~Fujimoto, \emph{Small Antennas: Miniaturization
  Techniques \& Applications}.\hskip 1em plus 0.5em minus 0.4em\relax
  McGraw-Hill, 2010.

\bibitem{Thal2006}
H.~L. Thal, ``New radiation {Q} limits for spherical wire antennas,''
  \emph{IEEE Trans. Antennas Propag.}, vol.~54, no.~10, pp. 2757--2763, Oct.
  2006.

\bibitem{McLean1996}
J.~S. McLean, ``A re-examination of the fundamental limits on the radiation
  {$Q$} of electrically small antennas,'' \emph{IEEE Trans. Antennas Propag.},
  vol.~44, no.~5, pp. 672--676, May 1996.

\bibitem{Kim2012}
O.~Kim, ``Minimum {Q} electrically small antennas,'' \emph{IEEE Trans. Antennas
  Propag.}, vol.~60, no.~8, pp. 3551--3558, Aug 2012.

\bibitem{Wheeler1947}
H.~A. Wheeler, ``Fundamental limitations of small antennas,'' \emph{Proc. IRE},
  vol.~35, no.~12, pp. 1479--1484, 1947.

\bibitem{Gustafsson+etal2015b}
M.~Gustafsson, D.~Tayli, and M.~Cismasu, \emph{Physical bounds of
  antennas}.\hskip 1em plus 0.5em minus 0.4em\relax Sprin\-ger-Verlag, 2015,
  pp. 1--32.

\bibitem{Gustafsson+etal2007a}
M.~Gustafsson, C.~Sohl, and G.~Kristensson, ``Physical limitations on antennas
  of arbitrary shape,'' \emph{Proc. R. Soc. A}, vol. 463, pp. 2589--2607, 2007.

\bibitem{Yaghjian+etal2013}
A.~D. Yaghjian, M.~Gustafsson, and B.~L.~G. Jonsson, ``Minimum {Q} for lossy
  and lossless electrically small dipole antennas,'' \emph{Progress In
  Electromagnetics Research}, vol. 143, pp. 641--673, 2013.

\bibitem{CapekJelinek_OptimalCompositionOfModalCurrentsQ}
M.~Capek and L.~Jelinek, ``Optimal composition of modal currents for minimal
  quality factor {Q},'' \emph{IEEE Trans. Antennas Propag.}, vol.~64, no.~12,
  pp. 5230--5242, 2016.

\bibitem{Sievenpiper+etal2012}
D.~F. Sievenpiper, D.~C. Dawson, M.~M. Jacob, T.~Kanar, S.~Kim, J.~Long, and
  R.~G. Quarfoth, ``Experimental validation of performance limits and design
  guidelines for small antennas,'' \emph{IEEE Trans. Antennas Propag.},
  vol.~60, no.~1, pp. 8--19, Jan 2012.

\bibitem{Thal2012}
H.~L. Thal, ``{Q Bounds for Arbitrary Small Antennas: A Circuit Approach},''
  \emph{IEEE Trans. Antennas Propag.}, vol.~60, no.~7, pp. 3120--3128, 2012.

\bibitem{Jonsson+Gustafsson2015}
B.~L.~G. Jonsson and M.~Gustafsson, ``Stored energies in electric and magnetic
  current densities for small antennas,'' \emph{Proc. R. Soc. A}, vol. 471, no.
  2176, p. 20140897, 2015.

\bibitem{Best2005}
S.~R. Best, ``Low {Q} electrically small linear and elliptical polarized
  spherical dipole antennas,'' \emph{IEEE Trans. Antennas Propag.}, vol.~53,
  no.~3, pp. 1047--1053, 2005.

\bibitem{JelinekCapek_OptimalCurrentsOnArbitrarilyShapedSurfaces}
L.~Jelinek and M.~Capek, ``Optimal currents on arbitrarily shaped surfaces,''
  \emph{IEEE Trans. Antennas Propag.}, vol.~65, no.~1, pp. 329--341, Jan. 2017.

\bibitem{GustafssonTayliEhrenborgEtAl_AntennaCurrentOptimizationUsingMatlabAndCVX}
\BIBentryALTinterwordspacing
M.~Gustafsson, D.~Tayli, C.~Ehrenborg, M.~Cismasu, and S.~Nordebo, ``Antenna
  current optimization using {MATLAB} and {CVX},'' \emph{{FERMAT}}, vol.~15,
  no.~5, pp. 1--29, May--June 2016. [Online]. Available:
  \url{http://www.e-fermat.org/articles/gustafsson-art-2016-vol15-may-jun-005/}
\BIBentrySTDinterwordspacing

\bibitem{Jonsson+Gustafsson2016}
B.~L.~G. Jonsson and M.~Gustafsson, ``Stored energies for electric and magnetic
  current densities,'' \emph{ArXiv e-print: 1604.08572}, pp. 1--25, 2016.

\bibitem{Fujimoto_Morishita_ModernSmallAntennas}
K.~Fujimoto and H.~Morishita, \emph{Modern Small Antennas}.\hskip 1em plus
  0.5em minus 0.4em\relax Cambridge University Press, 2013.

\bibitem{atom}
\BIBentryALTinterwordspacing
(2017) {A}ntenna {T}oolbox for {MATLAB} ({AToM}). Czech Technical University in
  Prague. [Online]. Available: \url{www.antennatoolbox.com}
\BIBentrySTDinterwordspacing

\bibitem{Meys_Diff_Method}
F.~J. R.~Meys, ``Measuring the impedance of balanced antennas by an s-parameter
  method.'' \emph{IEEE Antennas and Propagation}, vol.~40, no.~6, pp. 62--65,
  Dec. 1998.

\bibitem{RazaYangHussain_MeasurementOfRadEff_2012}
H.~Raza, J.~Yang, and A.~Hussain, ``Measurement of radiation efficiency of
  multiport antennas with feeding network corrections,'' \emph{IEEE Antennas
  and Wireless Propagation Letters}, vol.~11, pp. 89--92, 2012.

\bibitem{Pantoja_2003_PreFractal}
M.~F. Pantoja, F.~G. Ruiz, A.~R. Bretones, R.~G. Martin, J.~M. Gonzalez-Arbesu,
  J.~Romeu, and J.~M. Rius, ``{GA} design of wire pre-fractal antennas and
  comparison with other euclidean geometries,'' \emph{IEEE Antennas and
  Wireless Propagation Letters}, vol.~2, pp. 238--241, 2003.

\bibitem{BestMorrow_OnTheSignificanceOfCurrentVectorAlignmentInEstablishingTheResonantFrequencyOfSmallSpaceFillingWireAntennas}
S.~R. Best and J.~D. Morrow, ``On the significance of current vector alignment
  in establishing the resonant frequency of small space-filling wire
  antennas,'' \emph{IEEE Antennas Wireless Propag. Lett.}, vol.~2, pp.
  201--204, 2003.

\bibitem{Freescale1991}
A.~Wood and B.~Davidson, ``{RF} power device impedances: Practical
  considerations,'' Freescale Semiconductor, Inc., Tech. Rep. AN1526, 1991,
  rev. 0, 12/1991.

\bibitem{Marrocco_2008_ArtOfUHFRFID}
G.~Marrocco, ``The art of {UHF} {RFID} antenna design: impedance-matching and
  size-reduction techniques,'' \emph{IEEE Antennas and Propagation Magazine},
  vol.~50, no.~1, pp. 66--79, Feb 2008.

\bibitem{Pozar_MicrowaveEngineering}
D.~M. Pozar, \emph{Microwave Engineering}, 3rd~ed.\hskip 1em plus 0.5em minus
  0.4em\relax New York, NY: John Wiley \& Sons, 2005.

\bibitem{best2016optimizing}
S.~R. Best, ``Optimizing the receiving properties of electrically small {HF}
  antennas,'' \emph{URSI Radio Science Bulletin}, vol.~89, no.~4, pp. 13--29,
  2016.

\bibitem{Smith_1977_TAP}
G.~S. Smith, ``Efficiency of electrically small antennas combined with matching
  networks,'' \emph{IEEE Trans. Antennas Propag.}, vol.~25, pp. 369--373, 1977.

\bibitem{best2008small}
S.~R. Best, ``Small and fractal antennas,'' \emph{Modern antenna handbook}, pp.
  475--528, 2008.

\bibitem{nagaoka1909inductance}
H.~Nagaoka, ``The inductance coefficients of solenoids,'' \emph{Journal of the
  College of Science}, vol.~27, pp. 1--33, 1909, article 6.

\bibitem{GustafssonCapekSchab_Tradeoffs2017}
M.~Gustafsson, M.~Capek, and K.~Schab, ``\BIBforeignlanguage{eng}{Trade-off
  between antenna efficiency and {Q}-factor},'' Electromagnetic Theory
  Department of Electrical and Information Technology Lund University Sweden,
  Tech. Rep., 2019.

\bibitem{Palmier}
\BIBentryALTinterwordspacing
Palmier, \emph{Wikipedia, The Free Encyclopedia}. [Online]. Available:
  \url{https://en.wikipedia.org/wiki/Palmier}
\BIBentrySTDinterwordspacing

\bibitem{2018_Shahpari_TAP}
M.~Shahpari and D.~V. Thiel, ``Fundamental limitations for antenna radiation
  efficiency,'' \emph{IEEE Trans. Antennas Propag.}, vol.~66, no.~8, pp.
  3894--3901, Aug. 2018.

\bibitem{Harrington_EffectsOfAntennaSizeOnGainBWandEfficiency}
R.~F. Harrington, ``Effects of antenna size on gain, bandwidth, and
  efficiency,'' \emph{J. Nat. Bur. Stand.}, vol. 64-D, pp. 1--12, 1960.

\bibitem{UzsokySolymar_TheoryOfSuperDirectiveLinearArrays}
M.~Uzsoky and L.~Solym\'{a}r, ``Theory of super-directive linear arrays,''
  \emph{Acta Physica Academiae Scientiarum Hungaricae}, vol.~6, no.~2, pp.
  185--205, December 1956.

\bibitem{Harrington_AntennaExcitationForMaximumGain}
R.~F. Harrington, ``Antenna excitation for maximum gain,'' \emph{IEEE Trans.
  Antennas Propag.}, vol.~13, no.~6, pp. 896--903, Nov. 1965.

\bibitem{Jelinek2018TAPCost}
L.~Jelinek, K.~Schab, and M.~Capek, ``Radiation efficiency cost of resonance
  tuning,'' \emph{{IEEE} Transactions on Antennas and Propagation}, vol.~66,
  no.~12, pp. 6716--6723, December 2018.

\bibitem{Pfeiffer_FundamentalEfficiencyLimtisForESA}
C.~Pfeiffer, ``Fundamental efficiency limits for small metallic antennas,''
  \emph{IEEE Trans. Antennas Propag.}, vol.~65, pp. 1642--1650, 2017.

\bibitem{Thal2018}
H.~L. Thal, ``Radiation efficiency limits for elementary antenna shapes,''
  \emph{IEEE Trans. Antennas Propag.}, vol.~66, no.~5, pp. 2179--2187, May
  2018.

\bibitem{Gustafsson+Nordebo2013}
M.~Gustafsson and S.~Nordebo, ``Optimal antenna currents for {Q},
  superdirectivity, and radiation patterns using convex optimization,''
  \emph{IEEE Trans. Antennas Propag.}, vol.~61, no.~3, pp. 1109--1118, 2013.

\bibitem{Jonsson+etal2017b}
B.~L.~G. Jonsson, S.~Shi, F.~Ferrero, and L.~Lizzi, ``On methods to determine
  bounds on the {Q}-factor for a given directivity,'' \emph{IEEE Trans.
  Antennas Propag.}, vol.~65, no.~11, pp. 5686--5696, 2017.

\bibitem{Ferrero+etal2017}
\BIBentryALTinterwordspacing
F.~Ferrero, L.~Lizzi, B.~L.~G. Jonsson, and L.~Wang, ``A two-element parasitic
  antenna that approach the minimum {Q}-factor at a given directivity,''
  \emph{ArXiv e-prints, 1705.02281}, pp. 1--11, 2017. [Online]. Available:
  \url{http://arxiv.org/abs/1705.02281}
\BIBentrySTDinterwordspacing

\bibitem{Shi+etal2017}
\BIBentryALTinterwordspacing
S.~Shi, L.~Wang, and B.~L.~G. Jonsson, ``Antenna current optimization and
  realizations for far-field pattern shaping,'' \emph{ArXiv e-print,
  1711.09709v2}, 2018. [Online]. Available:
  \url{http://arxiv.org/abs/1711-09709v2}
\BIBentrySTDinterwordspacing

\bibitem{Yaghjian+etal2008}
A.~D. Yaghjian, T.~H. O'Donnell, E.~E. Altshuler, and S.~R. Best,
  ``Electrically small supergain end-fire arrays,'' \emph{Radio Science},
  vol.~43, no.~3, pp. 1--13, 2008.

\bibitem{Pigeon+etal2014}
M.~Pigeon, C.~Delaveaud, L.~Rudant, and K.~Belmkaddem, ``Miniature directive
  antennas,'' \emph{International Journal of Microwave and Wireless
  Technologies}, vol.~6, no.~1, pp. 45--50, 2014.

\bibitem{Ziolkowski+etal2013}
R.~W. Ziolkowski, M.-C. Tang, and N.~Zhu, ``An efficient, broad bandwidth, high
  directivity, electrically small antenna,'' \emph{Microwave and Optical
  Technology Letters}, vol.~55, no.~6, pp. 1430--1434, 2013.

\bibitem{Kim+etal2012}
O.~S. Kim, S.~Pivnenko, and O.~Breinbjerg, ``Superdirective magnetic dipole
  array as a first-order probe for spherical near-field antenna measurements,''
  \emph{IEEE Transactions on Antennas and Propagation}, vol.~60, no.~10, pp.
  4670--4676, Oct 2012.

\end{thebibliography}


\end{document}